\documentclass[12pt]{article}

\usepackage{amsmath,amssymb}
\usepackage{graphicx}
\usepackage{caption}
\usepackage{color}
\usepackage{dcolumn}
\usepackage{bm}
\usepackage[numbers,super,comma,sort&compress]{natbib}
\usepackage{url}

\definecolor{background-color}{gray}{0.98}

\DeclareMathOperator*{\argmin}{arg\,min}
\def\ddroit{{\rm d}}
\newcommand{\subsupi}[3]{#1_{\rm #2}^{\rm #3}}
\newcommand{\subi}[2]{#1_{\rm #2}}

\title{
Double hybrid density-functional theory using the Coulomb-attenuating method
}
\author{Yann Cornaton\thanks{Center for Theoretical and Computational
Chemistry, Departartment of Chemistry, Universitetet i Troms\o, Troms\o,
Norway}, Emmanuel Fromager\thanks{Laboratoire de Chimie Quantique, Institut de Chimie, CNRS/Universit\'{e} de Strasbourg, Strasbourg, France}}

\begin{document}

\maketitle

\begin{abstract}
A double hybrid
approximation using the Coulomb-attenuating method (CAM-DH) is derived
within range-separated
density-functional perturbation theory, in the spirit of a recent work 
by Cornaton {\it et al.} [Phys. Rev. A
88, 022516 (2013)]. The energy expression recovered through second order is
linear in the parameters $\alpha$ and $\beta$ that control the Coulomb
attenuation. The method has been tested within the local density
approximation on a small test set consisting
of rare-gas and alkaline-earth-metal dimers as well as diatomics with
single, double and triple bonds. In this context, 
the semi-empirical $\alpha=0.19$ and
$\beta=0.46$ parameters, that were optimized for the hybrid
CAM-B3LYP functional, do not provide accurate interaction and total
energies. 
Using semi-local functionals 
with density scaling, that was neglected in this work, may lead to different
conclusions. Calibration studies on a larger test set would be necessary
at this point. This is left for future work. Finally, we propose as a perspective an alternative CAM-DH
approach that relies on the perturbation expansion of a partially
long-range interacting wavefunction. In this case the energy is not
linear anymore in $\alpha$ and $\beta$. Work is in progress in this
direction. 
\end{abstract}

\clearpage


  \makeatletter
  \renewcommand\@biblabel[1]{#1.}
  \makeatother

\bibliographystyle{apsrev}

\renewcommand{\baselinestretch}{1.5}
\normalsize

\clearpage

\section{\sffamily \Large INTRODUCTION} 

The combination of density-functional theory (DFT) with second-order M\o
ller--Plesset (MP2)
perturbation theory can be achieved rigorously when splitting the
electron-electron repulsion into two complementary
contributions~\cite{Angyan2005PRA,Goll2005PCCP,Fromager2008PRA,2blehybrids_Julien,2blehybrids_Fromager2011JCP,AC_2blehybrids_Yann}. Note that, even though we focus here on MP2,
various correlated methods have been merged with DFT along those lines
(see Ref.~\cite{Cornaton2013PRA} and the references therein). The resulting MP2-DFT energy expressions are usually
referred to as double hybrid approximations. So far two separations of
the two-electron interaction have
been investigated: one is simply linear~\cite{2blehybrids_Julien} and
the other one is based on the range of the interaction, thus
leading to the so-called long-range/short-range
separation~\cite{Savin1996Book}.\\

These separations have also been used in conventional (single determinantal)
hybrid DFT for the purpose of improving 
the description of the exchange energy. While the linear separation
underlies popular hybrid functionals such as 
the Becke
three-parameter Lee-Yang-Parr functional (B3LYP)~\cite{refb3lyp},
standard long-range-corrected hybrid DFT (LC-DFT)~\cite{hiraomu} uses 
the range
separation based on the error function. The combination of the two
latter approaches lead to the Coulomb-attenuated method based
on the B3LYP functional (CAM-B3LYP)~\cite{Yanai2004CPL}. While
preserving the accuracy of B3LYP for ground-state properties, CAM-B3LYP
became popular for the computation of charge-transfer excitations within
time-dependent DFT~\cite{peach2008}.\\

We explore in this work rigorous double hybrid extensions for
CAM-B3LYP with the purpose of improving both exchange and correlation
ground-state energies. The paper is organized as follows: In
Sec.~\ref{sec:theory} the theory underlying Coulomb-attenuating double hybrid DFT
is presented. The latter will be based on the perturbation expansion of
a fully long-range interacting wavefunction. Computational details are
then given in Sec.~\ref{sec:comp_details} 
and results obtained on a small
test set, consisting of rare-gas and
alkaline-earth-metal dimers as well as diatomics with single, double and
triple bonds, are discussed in Sec.~\ref{sec:discussion}. As a perspective, we finally propose in
Sec.~\ref{CAM-DH_alternative} an alternative formulation that relies on
a partially long-range-interacting wavefunction. Conclusions are given
in Sec.~\ref{sec:conclusions}.

\section{\sffamily \Large THEORY}\label{sec:theory}

In this section we present the theory underlying the 
construction of CAM-DH
approximations. It is organized as follows: For pedagogical purposes,
standard hybrid LC-DFT and 
range-separated
double hybrid DFT are introduced in Sec.~\ref{subsec:h_and_rsdh}.
We then discuss the multi-determinantal extension of standard
hybrid CAM-DFT in Sec.~\ref{subsec:md-camdft}. In
Sec.~\ref{subsec:2bleAC} we consider a double adiabatic connection and apply scaling relations in order to derive implementable 
expressions for complement 
density-functional correlation energies. A CAM-DH energy expression is 
finally
derived in Sec.~\ref{CAM-DH_approx}.  

\subsection{\sffamily \Large 
Hybrid and double hybrid DFT based on range separation
}\label{subsec:h_and_rsdh}

\subsubsection{Long-range corrected hybrid DFT}

According to the Hohenberg--Kohn theorems~\cite{Hohenberg1964PR}, the exact ground-state energy
of an electronic system can be expressed as
\begin{align}\label{VarprincipleHK}
E = \underset{n}{\rm min}\left\{ F[n]+\int {\rm d}{\bf r}\,v_{\rm
ne}({\bf r})\,n({\bf r}) \right\}, 
\end{align}
where $v_{\rm ne}({\bf r})$ is the nuclear potential and $F[n]$ denotes
the universal Levy--Lieb (LL) functional~\cite{LevyF,LiebF} 
\begin{align}\label{univfunc}
F[n] &= 
\underset{\Psi\rightarrow n}{\rm min}\langle \Psi\vert
\hat{T}+\hat{W}_{\rm ee}\vert\Psi\rangle
.
\end{align}
$\hat{T}$ is the kinetic energy operator and
$\hat{W}_{\rm ee}$ denotes the regular two-electron interaction
operator with $w_{\rm ee}(r_{12})=1/r_{12}$. The minimization in Eq.~(\ref{univfunc}) is restricted to wavefunctions
with density $n$.\\

In standard hybrid LC-DFT~\cite{hiraomu}, the following
partitioning of the LL functional is used 
\begin{align}\label{LL_parti_LChyb}
F[n]=
\Bigg(
\underset{\Phi\rightarrow n}{\rm min}\langle \Phi\vert
\hat{T}+\hat{W}^{\rm lr,\mu}_{\rm ee}\vert\Phi\rangle
\Bigg)+
E^{\rm sr,\mu}_{\rm H}[n]+
E^{\rm sr,\mu}_{\rm x}[n]
+{U}^\mu_{\rm c}[n
],
\end{align}
where the minimization in the first term on the right-hand side of
Eq.~(\ref{LL_parti_LChyb}) is restricted to single determinants $\Phi$ with density $n$, 
$\hat{W}^{\rm lr,\mu}_{\rm ee}$
is the long-range electron-electron interaction operator defined by
$w^{\rm lr,\mu}_{\rm ee}(r_{12})={\rm erf}(\mu r_{12})/r_{12}$ and
$E^{\rm sr,\mu}_{\rm H}[n]=1/2\int \int \ddroit {\mathbf{r}_1}\ddroit
{\mathbf{r}_2}n({\mathbf{r}_1})n({\mathbf{r}_2})\,w^{\rm sr,\mu}_{\rm
ee}(r_{12})$ denotes the short-range Hartree density functional with
$w^{\rm sr,\mu}_{\rm ee}(r_{12})={\rm erfc}(\mu r_{12})/r_{12}$. 
In this scheme the
range separation is controlled by the $\mu$ parameter. 
Note that, for $\mu=0$, the long-range interaction equals zero and the
short-range interaction reduces to the regular two-electron interaction
$1/r_{12}$, thus leading to the standard Kohn--Sham (KS)
decomposition~\cite{kstheo} 
\begin{align}\label{LL_parti_KS}
F[n]=T_{\rm s}[n]
+E_{\rm H}[n]+
E_{\rm x}[n]
+{E}_{\rm c}[n
],
\end{align}
where $T_{\rm s}[n]=
\langle\Phi^{\rm KS}[n] \vert
\hat{T}\vert\Phi^{\rm KS}[n]\rangle
$ is the non-interacting kinetic energy functional and $\Phi^{\rm
KS}[n]$ denotes the KS determinant with density $n$. The latter enables
to define the exact short-range exchange energy in
Eq.~(\ref{LL_parti_LChyb}) as 
\begin{align}
E^{\rm sr,\mu}_{\rm
x}[n]=
\langle\Phi^{\rm KS}[n] \vert\hat{W}^{\rm sr,\mu}_{\rm
ee}\vert\Phi^{\rm KS}[n]\rangle
-E^{\rm sr,\mu}_{\rm H}[n],
\end{align}
which gives the following expression for
the {\it exact} complement correlation functional 
\begin{align}\label{Ec_LChyb}
{U}^\mu_{\rm c}[n]={E}_{\rm c}[n]+
\langle \Phi^{\rm KS}[n]\vert
\hat{T}+\hat{W}^{\rm lr,\mu}_{\rm ee}\vert\Phi^{\rm KS}[n]\rangle
-
\underset{\Phi\rightarrow n}{\rm min}\langle \Phi\vert
\hat{T}+\hat{W}^{\rm lr,\mu}_{\rm ee}\vert\Phi\rangle.
\end{align}
Combining Eq.~(\ref{VarprincipleHK}) with
Eq.~(\ref{LL_parti_LChyb}) leads to the following 
expression for the exact ground-state energy 
\begin{eqnarray}\label{E_LChyb_min}
E&=&\min_\Phi\left\{\langle\Phi|\hat{T}+\hat{V}_{\rm ne}+\hat{W}^{\rm
lr,\mu}_{\rm ee}|\Phi\rangle+
E^{\rm sr,\mu}_{\rm H}[n_{\Phi}]+
E^{\rm sr,\mu}_{\rm x}[n_{\Phi}]
+{U}^\mu_{\rm c}[n_{\Phi}
]
\right\}
,
\end{eqnarray}
where the nuclear potential operator equals $\hat{V}_{\rm ne}=
\int \ddroit{\bf r}\,v_{\rm ne}({\bf r})\,\hat{n}({\bf r})$ and
$\hat{n}({\bf r})=\sum_{\sigma=\alpha,\beta}\hat{\Psi}_\sigma^\dagger({\bf
r})\hat{\Psi}_\sigma({\bf r})$ is the density operator written in
second quantized form. 
Note that, in practical calculations, ${U}^\mu_{\rm c}[n]$ is simply replaced by
the regular correlation functional ${E}_{\rm
c}[n]$ since the last two terms on the right-hand side of Eq.~(\ref{Ec_LChyb})
are expected to be relatively close, like in conventional hybrid
DFT~\cite{BECKE:1993p1788}.  
Let us stress that in 
hybrid LC-DFT, range separation is only used for the exchange energy. The correlation energy is, like in
KS-DFT, described by a density functional.
Consequently one single determinant is sufficient for computing the
ground-state energy. The latter becomes \emph{exact} when
both exact short-range exchange and complement correlation
functionals are used.
\\ 

\subsubsection{Multi-determinant range-separated
DFT}\label{subsubsec:mdrsdft}
In order to improve the description of the long-range 
correlation energy in approximate LC-DFT schemes, Savin has proposed~\cite{Savin1996Book} a
multi-determinantal extension of Eq.~(\ref{E_LChyb_min}) based on the
following decomposition of the LL functional:
\begin{eqnarray}
\label{LL_parti_srdft}
F[n]&=&
\Bigg(
\underset{\Psi\rightarrow n}{\rm min}
\langle \Psi\vert
\hat{T}+\hat{W}^{\rm lr,\mu}_{\rm ee}\vert\Psi\rangle
\Bigg)
+
E^{\rm sr,\mu}_{\rm Hxc}[n]
\nonumber\\
&=&
\langle \Psi^\mu[n]\vert
\hat{T}+\hat{W}^{\rm lr,\mu}_{\rm ee}\vert\Psi^\mu[n]\rangle
+E^{\rm sr,\mu}_{\rm Hxc}[n]
,
\end{eqnarray}
where the complement $\mu$-dependent short-range Hartree-exchange-correlation
(srHxc) density-functional energy is denoted $E^{\rm sr,\mu}_{\rm
Hxc}[n]$. Note that, in contrast to hybrid LC-DFT (see
Eq.~(\ref{LL_parti_LChyb})), the minimization in
the first term on the right-hand side of Eq.~(\ref{LL_parti_srdft}) is
{\it not} restricted to single determinants.
Consequently, the minimizing wavefunction $\Psi^\mu[n]$ with density $n$
is \emph{multi-determinantal}. In other
words, purely long-range correlation effects are now treated explicitly,
in wavefunction theory. Note that the LL universal functional expression in
Eq.~(\ref{univfunc}) 
is recovered from Eq.~(\ref{LL_parti_srdft}) in the $\mu\rightarrow +\infty$ limit.
The exact short-range exchange energy is usually defined, like in hybrid
LC-DFT, from the KS determinant thus leading to the
following expression for the srHxc energy:  
\begin{eqnarray} \label{srDFTfunhxcdef}
E^{\rm sr,\mu}_{\rm Hxc}[n]&=&E^{\rm sr,\mu}_{\rm H}[n]+E^{\rm sr,\mu}_{\rm x}[n]
+
E^{\rm sr,\mu}_{\rm c}[n]
\nonumber\\
&=&
\langle \Phi^{\rm KS}[n]\vert
\hat{W}^{\rm sr,\mu}_{\rm ee}\vert\Phi^{\rm KS}[n]\rangle
+
E^{\rm sr,\mu}_{\rm c}[n]
,
\end{eqnarray}
where, according to Eqs.~(\ref{LL_parti_KS}) and (\ref{LL_parti_srdft}),
the complement short-range correlation energy can be expressed
as 
\begin{eqnarray} \label{srDFTfuncdef}
E^{\rm sr,\mu}_{\rm c}[n]&=&
T_{\rm s}[n]+E_{\rm H}[n]-E^{\rm sr,\mu}_{\rm H}[n]
+E_{\rm x}[n]-E^{\rm sr,\mu}_{\rm x}[n]
+E_{\rm c}[n]
\nonumber\\
&&-
\Bigg(
\underset{\Psi\rightarrow n}{\rm min}
\langle \Psi\vert
\hat{T}+\hat{W}^{\rm lr,\mu}_{\rm ee}\vert\Psi\rangle
\Bigg).
\end{eqnarray}
By using the KS decomposition of the long-range interacting LL
functional,
\begin{eqnarray} \label{KSdecomp_lrLL}
\underset{\Psi\rightarrow n}{\rm min}
\langle \Psi\vert
\hat{T}+\hat{W}^{\rm lr,\mu}_{\rm ee}\vert\Psi\rangle
=
T_{\rm s}[n]+
\langle \Phi^{\rm KS}[n]\vert
\hat{W}^{\rm lr,\mu}_{\rm ee}\vert\Phi^{\rm KS}[n]\rangle
+E^{\rm lr,\mu}_{\rm c}[n],
\end{eqnarray}
where $E^{\rm lr,\mu}_{\rm c}[n]$ denotes the purely long-range
density-functional correlation energy, we obtain the compact expression
\begin{eqnarray} \label{srDFTfuncdef_final}
E^{\rm sr,\mu}_{\rm c}[n]=
E_{\rm c}[n]-E^{\rm lr,\mu}_{\rm c}[n].
\end{eqnarray}
Local density approximations (LDA) to the short-range correlation functional have
been developed along those lines when substituting 
$w^{\rm lr,\mu}_{\rm
ee}(r_{12})$ for $1/r_{12}$ 
in the uniform electron
gas model~\cite{Toulouse2004IJQC,Paziani2006PRB}.\\   

Returning to the exact theory, combining Eq.~(\ref{VarprincipleHK}) with
Eq.~(\ref{LL_parti_srdft}) leads to 
\begin{eqnarray}\label{E_srdft_min}
E&=&\min_\Psi\left\{\langle\Psi|\hat{T}+\hat{V}_{\rm ne}+\hat{W}^{\rm lr,\mu}_{\rm ee}|\Psi\rangle+E^{\rm sr,\mu}_{\rm Hxc}[n_{\Psi}]\right\}
\nonumber
\\
&=&\langle\Psi^\mu|\hat{T}+\hat{V}_{\rm ne}+\hat{W}^{\rm lr,\mu}_{\rm ee}|\Psi^\mu\rangle+E^{\rm sr,\mu}_{\rm Hxc}[n_{\Psi^\mu}]
,
\end{eqnarray}
where the exact minimizing wavefunction $\Psi^\mu$ is
multi-determinantal due to the explicit description of the long-range
interaction. As discussed further in Sec.~\ref{subsubsec:RSDHs}, applying MP2 in this context
leads to the formulation of range-separated double hybrid approximations.\\

We should finally stress that the decomposition in
Eq.~(\ref{srDFTfunhxcdef}) is \emph{not} unique. As mentioned in
previous
works~\cite{Toulouse2005TCA,Gori-Giori2006PRA,Gori-Giorgi2009IJQC} 
, it seems natural in this context to use the
multi-determinantal (md) long-range interacting wavefunction $\Psi^\mu[n]$
with density $n$ rather than the KS determinant for the separation of
short-range exchange and correlation energies:   
\begin{eqnarray}\label{eq:mdHx_def}
E^{\rm sr,\mu}_{\rm Hxc}[n]
=\langle\Psi^\mu[n]|\hat{W}^{\rm sr,\mu}_{\rm
ee}|\Psi^\mu[n]\rangle+
E^{\rm sr,\mu}_{\rm c,md}[n].
\end{eqnarray}
An adapted complement short-range correlation functional, denoted $E^{\rm sr,\mu}_{\rm
c,md}[n]$, must be used rather than the usual short-range correlation
functional $E^{\rm sr,\mu}_{\rm c}[n]$ in order to recover the same
srHxc energy from both decompositions:
\begin{eqnarray}\label{eq:mdc_expression}
E^{\rm sr,\mu}_{\rm c,md}[n]
=E^{\rm sr,\mu}_{\rm c}[n]
+\langle \Phi^{\rm KS}[n]\vert
\hat{W}^{\rm sr,\mu}_{\rm ee}\vert\Phi^{\rm KS}[n]\rangle
-\langle\Psi^\mu[n]|\hat{W}^{\rm sr,\mu}_{\rm
ee}|\Psi^\mu[n]\rangle
.
\end{eqnarray}
In this work, the LDA-type short-range md correlation functional
of Paziani {\it et
al.}~\cite{Paziani2006PRB} will be used. 

Returning to the exact
theory, since~\cite{Cornaton2013PRA} $\Psi^\mu[n_{\Psi^\mu}]=\Psi^\mu$,  
combining Eq.~(\ref{E_srdft_min}) with Eq.~(\ref{eq:mdHx_def}) leads to the 
alternative range-separated expression for the ground-state energy, 
\begin{eqnarray}\label{E_srDFT_md}
E=\langle\Psi^\mu|\hat{T}+\hat{V}_{\rm ne}+\hat{W}_{\rm
ee}|\Psi^\mu\rangle+E^{\rm sr,\mu}_{\rm c,md}[n_{\Psi^\mu}]
,
\end{eqnarray}
where long- and short-range interactions have been recombined.  

\subsubsection{Range-separated double hybrids}\label{subsubsec:RSDHs}

As shown in
Refs.~~\cite{Angyan2005PRA,Fromager2008PRA,pra_MBPTn-srdft_janos,Fromager2011JCP,Cornaton2013PRA}, rigorous range-separated double hybrid (RSDH)
energy expressions can be derived from Eqs.~(\ref{E_srdft_min}) and
(\ref{E_srDFT_md}) by expanding the multi-determinantal wavefunction
$\Psi^\mu$ in a self-consistent MP2-type density-functional perturbation
theory. Key ideas are the following: By analogy with the regular
Hartree--Fock (HF) approximation, the minimization in
Eq.~(\ref{E_srdft_min}) is first restricted to single determinantal
wavefunctions $\Phi$, 
\begin{eqnarray}
E^{\rm srDFT}_{\rm HF}&=&\underset{\Phi}{\rm min}\left\{\!\langle \Phi \vert
\hat{T}+\hat{W}^{\rm lr,\mu}_{\rm ee}+\hat{V}_{\rm
ne}\vert\Phi\rangle+E^{\rm sr,\mu}_{\rm Hxc}[n_{\Phi}] \right\}
\nonumber\\
&=&
\langle\Phi_0^\mu|\hat{T}+\subsupi{\hat{W}}{ee}{lr,\mu}+\subi{\hat{V}}{ne}
|\Phi_0^\mu\rangle
+\subsupi{E}{Hxc}{sr,\mu}[n_{\Phi_0^\mu}],
\end{eqnarray}
thus defining the HF-short-range DFT (HF-srDFT) approximation.
The minimizing determinant $\Phi^\mu_0$, referred to as HF-srDFT
determinant, fulfills the following HF-type
equation: 
\begin{align}
\Bigg(\hat{T}+\subsupi{\hat{U}}{HF}{lr,\mu}+ \hat{V}_{\rm ne}
+\int \ddroit \mathbf{r}\,\frac{\delta E^{\rm sr,\mu}_{\rm
Hxc}}{\delta n(\mathbf{r})}[n_{\Phi^\mu_0}]\,\hat{n}(\mathbf{r})
\Bigg)\vert\Phi^\mu_0\rangle=\mathcal{E}_0^\mu\vert\Phi^\mu_0\rangle,
\end{align}
where $\subsupi{\hat{U}}{HF}{lr,\mu}$ is the long-range analogue of the
HF potential operator calculated with the occupied HF-srDFT orbitals.
We then 
introduce a perturbation strength $\epsilon$ and define the auxiliary
energy~\cite{Angyan2005PRA}
\begin{eqnarray}\label{Eepsilon_var-pt}
E^{\epsilon,\mu} &=& \underset{\Psi}{\rm min}\Big\{ 
\langle \Psi\vert
\hat{T} + \hat{V}_{\rm ne} +
(1-\epsilon)\subsupi{\hat{U}}{HF}{lr,\mu}
+
\epsilon\subsupi{\hat{W}}{ee}{lr,\mu}
\vert\Psi\rangle
+E^{\rm sr,\mu}_{\rm Hxc}[n_{\Psi}]\Big\}.  
\end{eqnarray}
Note that, according to Eq.~(\ref{E_srdft_min}), the \emph{exact} ground-state
energy is recovered when $\epsilon=1$. 
As discussed in details in
Refs.~\cite{Angyan2005PRA,Fromager2008PRA,pra_MBPTn-srdft_janos,Fromager2011JCP}, the minimizing wavefunction $\Psi^{\epsilon,\mu}$ in
Eq.~(\ref{Eepsilon_var-pt}) and its density $n_{\Psi^{\epsilon,\mu}}$ can be
expanded through second order in the long-range fluctuation potential
$\subsupi{\hat{W}}{ee}{lr,\mu}-\subsupi{\hat{U}}{HF}{lr,\mu}$ as follows, 
\begin{eqnarray}\label{wf-pertalpha}
|\Psi^{\epsilon,\mu}\rangle&=&|\Phi_0^\mu\rangle+\epsilon|\Psi^{\rm
(1)lr,\mu}\rangle+\epsilon^2|\Psi^{\rm
(2)\mu}\rangle+\mathcal{O}(\epsilon^3),
\end{eqnarray}
where the first-order contribution is the long-range analogue of the
MP1 wavefunction correction, and
\begin{eqnarray}\label{density-pert}
n_{\Psi^{\epsilon,\mu}}({\bf r})&=&n_{\Phi^\mu_0}({\bf r})+\epsilon^2\delta n^{(2)\mu}({\bf
r})+\mathcal{O}(\epsilon^3).
\end{eqnarray}
Since the density remains unchanged through first order,
the auxiliary energy is simply 
expanded through second order as~\cite{Angyan2005PRA,Fromager2008PRA}
\begin{eqnarray}\label{ptepsilon_second}\begin{array}{l}
\displaystyle 
{E}^{\epsilon,\mu}
={E}^{(0)\mu}+\epsilon{E}^{(1)\mu}+\epsilon^2{E}_{\rm MP}^{(2)\rm lr,\mu}+\mathcal{O}(\epsilon^3),
\end{array}
\end{eqnarray}
where, when considering the $\epsilon=1$ limit, the HF-srDFT energy is
recovered through first order, 
\begin{eqnarray}\label{HF-srDFT-eq2}
{E}^{(0)\mu}+{E}^{(1)\mu}
=\subsupi{E}{\tiny HF}{\tiny srDFT},
\end{eqnarray}
and the second-order correction to the energy is the purely long-range
MP2 correlation energy calculated with HF-srDFT orbitals and orbital
energies. The MP2-srDFT approximation is obtained by truncating the
perturbation expansion through second order, thus leading to the
following energy expression 
\begin{eqnarray}\label{mp2srdft-ener_exp}
E^{\rm srDFT}_{\rm MP2}
&=&
\langle\Phi_0^\mu|\hat{T}+\subi{\hat{V}}{ne}
|\Phi_0^\mu\rangle+E_{\rm H}[n_{\Phi_0^\mu}]+
E^{\rm HF}_{\rm x}[\Phi_0^\mu]
-
E^{\rm HF, sr, \mu}_{\rm x}[\Phi_0^\mu]
\nonumber\\
&&+
E^{\rm
sr,\mu}_{\rm x}[n_{\Phi_0^\mu}]
+
E^{(2)\rm lr,\mu}_{\rm MP}
+E^{\rm sr,\mu}_{\rm c}[n_{\Phi_0^\mu}]
,
\end{eqnarray}
where $E^{\rm HF}_{\rm x}[\Phi_0^\mu]$ and $E^{\rm HF, sr, \mu}_{\rm
x}[\Phi_0^\mu]$ are the regular (full-range) and short-range HF exchange
energies, respectively, both obtained from the HF-srDFT determinant.
Eq.~(\ref{mp2srdft-ener_exp}) defines a RSDH approximation where
the exchange and correlation energies are 
\begin{eqnarray}\label{Ex_mp2srdft}
E^{\rm srDFT}_{\rm x,MP2}
&=&
E^{\rm HF}_{\rm x}[\Phi]
-
E^{\rm HF, sr, \mu}_{\rm x}[\Phi]
+
E^{\rm
sr,\mu}_{\rm x}[n],
\end{eqnarray}
and 
\begin{eqnarray}\label{Ec_mp2srdft}
E^{\rm srDFT}_{\rm c,MP2}
&=&
E^{(2)\rm lr,\mu}_{\rm MP}
+E^{\rm sr,\mu}_{\rm c}[n]
,
\end{eqnarray}
$\Phi$ and $n$ being shorthand notations for the HF-srDFT determinant and its density,
respectively.\\

Finally, as shown by Cornaton {\it et al.}~\cite{Cornaton2013PRA}, combining the
wavefunction expansion in Eq.~(\ref{wf-pertalpha}) with the alternative
energy expression in Eq.~(\ref{E_srDFT_md}) leads to another type of
RSDH approximation that involves, in the
computation of the energy through first order,
{\it full-range} integrals only. For that reason, the method was referred to
as RSDHf in Ref.~\cite{Cornaton2013PRA} When second-order corrections to the density are
neglected, the energy equals through second order~\cite{Cornaton2013PRA}
\begin{eqnarray}\label{rsdhf-ener_exp}
E_{\rm RSDHf}
&=&
\langle\Phi_0^\mu|\hat{T}+\subi{\hat{V}}{ne}
|\Phi_0^\mu\rangle+E_{\rm H}[n_{\Phi_0^\mu}]+
E^{\rm HF}_{\rm x}[\Phi_0^\mu]
\nonumber\\
&&
+
E^{(2)\rm lr,\mu}_{\rm MP}
+E^{(2)\rm lr-sr,\mu}_{\rm MP}
+E^{\rm sr,\mu}_{\rm c, md}[n_{\Phi_0^\mu}]
,
\end{eqnarray}
where 
$E^{(2)\rm lr-sr,\mu}_{\rm MP}$
denotes the MP2 coupling term between 
long-range and short-range correlations
calculated with HF-srDFT orbitals and orbital
energies. The corresponding exchange and correlation energies are
\begin{eqnarray}\label{rsdhf-xener_exp}
E_{\rm x, RSDHf}
=
E^{\rm HF}_{\rm x}[\Phi]
,
\end{eqnarray}
and
\begin{eqnarray}\label{rsdhf-cener_exp}
E_{\rm c, RSDHf}
=
E^{(2)\rm lr,\mu}_{\rm MP}
+E^{(2)\rm lr-sr,\mu}_{\rm MP}
+E^{\rm sr,\mu}_{\rm c, md}[n]
,
\end{eqnarray}
respectively, where the same shorthand notations as in
Eqs.~(\ref{Ex_mp2srdft}) and (\ref{Ec_mp2srdft}) are used.

\subsection{Multi-determinant DFT based on the Coulomb-attenuating
method}\label{subsec:md-camdft}

\subsubsection{Coulomb-attenuating hybrid DFT}

The partitioning of the universal LL functional underlying standard
hybrid CAM-DFT (whose most popular approximate formulation is
CAM-B3LYP~\cite{Yanai2004CPL}) is obtained from
Eq.~(\ref{LL_parti_LChyb}) by substituting
the $\alpha,\beta$-dependent attenuated interaction 
for the purely long-range one, 
\begin{align}\label{LC_to_cam}
w^{\rm lr,\mu}_{\rm ee}(r_{12})\rightarrow
\alpha w_{\rm ee}(r_{12})+\beta w^{\rm lr,\mu}_{\rm
ee}(r_{12}),
\end{align}
with  
the relations $0\leq \alpha+\beta\leq 1$, $0\leq\alpha\leq 1$ and
$0\leq\beta\leq1$, 
thus leading to
\begin{eqnarray}\label{LL_parti_camhyb}
F[n]&=&
\Bigg(
\underset{\Phi\rightarrow n}{\rm min}\langle \Phi\vert
\hat{T}+\alpha\hat{W}_{\rm ee}+
\beta\hat{W}^{\rm lr,\mu}_{\rm ee}
\vert\Phi\rangle
\Bigg)+
\big(1-\alpha-\beta\big)E_{\rm H}[n]
+\beta E^{\rm sr,\mu}_{\rm H}[n]
\nonumber
\\
&&
+\big(1-\alpha-\beta\big)E_{\rm x}[n]
+
\beta E^{\rm sr,\mu}_{\rm x}[n]
+{U}^{\mu,\alpha,\beta}_{\rm c}[n
].
\end{eqnarray}
According to the KS decomposition in
Eq.~(\ref{LL_parti_KS}), the \emph{exact} complement correlation functional can
be expressed as
\begin{eqnarray}\label{LL_corrfun_camhyb}
{U}^{\mu,\alpha,\beta}_{\rm c}[n]&=&
E_{\rm c}[n]+
\langle \Phi^{\rm KS}[n]\vert
\hat{T}+\alpha\hat{W}_{\rm ee}+
\beta\hat{W}^{\rm lr,\mu}_{\rm ee}
\vert\Phi^{\rm KS}[n]\rangle
\nonumber
\\
&&-
\Bigg(
\underset{
\Phi\rightarrow n}{\rm min}\langle \Phi\vert
\hat{T}+
\alpha\hat{W}_{\rm ee}+
\beta\hat{W}^{\rm lr,\mu}_{\rm ee}
\vert\Phi\rangle
\Bigg)
.
\end{eqnarray}
Combining Eq.~(\ref{VarprincipleHK}) with Eq.~(\ref{LL_parti_camhyb})
leads to the exact hybrid CAM-DFT energy expression
\begin{eqnarray}\label{E_camhyb_min}
E&=&\min_\Phi\Bigg\{\langle\Phi|\hat{T}+\hat{V}_{\rm ne}+
\alpha\hat{W}_{\rm ee}+
\beta\hat{W}^{\rm lr,\mu}_{\rm ee}
|\Phi\rangle+
\big(1-\alpha-\beta\big)E_{\rm H}[n_{\Phi}]
+\beta E^{\rm sr,\mu}_{\rm H}[n_{\Phi}]
\nonumber
\\
&&
+\big(1-\alpha-\beta\big)E_{\rm x}[n_{\Phi}]
+
\beta E^{\rm sr,\mu}_{\rm x}[n_{\Phi}]
+{U}^{\mu,\alpha,\beta}_{\rm c}[n_{\Phi}
]
\Bigg\}
.
\end{eqnarray}
Note that, in practical calculations, the stantard correlation
functional $E_{\rm c}[n]$ is used for ${U}^{\mu,\alpha,\beta}_{\rm
c}[n]$~\cite{Yanai2004CPL}. Thus it is assumed that the last two terms on the right-hand side of
Eq.~(\ref{LL_corrfun_camhyb}) compensate. Let us stress that, in hybrid
CAM-DFT, the CAM is used for the
exchange energy only. The correlation energy is, like in hybrid LC-DFT or
KS-DFT, described by a density functional.
Consequently one single determinant is sufficient for computing the
ground-state energy. The latter becomes \emph{exact} when
both exact short-range and regular (full-range) exchange energy density
functionals are used in conjunction with the exact complement correlation
functional.

\subsubsection{Multi-determinantal extensions}

We discuss in this section the multi-determinantal extension of hybrid
CAM-DFT. In the light of Sec.~\ref{subsubsec:mdrsdft}, the most natural way to
proceed consists in extending the minimization in the first term on the
right-hand side of 
Eq.~(\ref{LL_parti_camhyb}) to
multi-determinantal wavefunctions with density $n$. A complement correlation
functional, depending on both $\mu$, $\alpha$ and $\beta$, should then
be constructed so that the universal LL functional is
recovered from the new partitioning. The resulting exact expression for
the ground-state energy would then 
be formally identical to the range-separated one in
Eq.~(\ref{E_srdft_min}). The only
difference would come from the substitution in Eq.~(\ref{LC_to_cam}).  
Applying MP2 in this context would provide a CAM-DH energy expression.
Let us stress that such a CAM-DH is not expected to converge as fast as
MP2-srDFT and RSDHf with respect to the basis set~\cite{Cornaton2013PRA}
simply because, unlike the purely long-range interaction, the
Coulomb-attenuated interaction has a singularity at $r_{12}=0$. Of course,
using the CAM makes the electronic cusp condition 
weaker, since at short range the regular interaction $1/r_{12}$ is scaled by $\alpha$,
but still the singularity remains. In connection to this, basis set superposition errors
(BSSE) are also expected to be larger relative to
MP2-srDFT and RSDHf.\\   

We choose here not to explore further such a
CAM-DH scheme. We rather propose to keep the purely long-range MP2
wavefunction expansion underlying both MP2-srDFT and RSDHf while
introducing Coulomb attenuation into the energy expression. This can
be achieved rigorously by using the 
following decomposition of the srHxc density-functional
energy,
\begin{eqnarray}\label{mdEXX_alpha_beta}
E^{\rm sr, \mu}_{\rm
Hxc}[n]=\langle\Psi^\mu[n]\vert(\beta-1)\hat{W}^{\rm lr,\mu}_{\rm
ee}+\alpha\hat{W}_{\rm ee}|\Psi^\mu[n]\rangle+\bar{E}^{\mu,\alpha,\beta}_{\rm Hxc}[n]
,
\end{eqnarray}
where the complement three-parameter density functional
$\bar{E}_{\rm Hxc}^{\rm \mu,\alpha,\beta}[n]$ is such that the exact
($\mu$-dependent only) srHxc
energy is obtained for any values of $\alpha$ and $\beta$.
Since~\cite{Cornaton2013PRA} $\Psi^\mu[n_{\Psi^\mu}]=\Psi^\mu$, the exact ground-state energy
expression in Eq.~(\ref{E_srdft_min}) becomes with the partitioning
in Eq.~(\ref{mdEXX_alpha_beta}),
\begin{eqnarray}
E=\langle\Psi^\mu|\hat{T}+\hat{V}_{\rm ne}+\beta\hat{W}_{\rm ee}^{\rm lr, \mu}+\alpha\hat{W}_{\rm ee}|\Psi^\mu\rangle+\bar{E}_{\rm Hxc}^{\mu,\alpha,\beta}[n_{\Psi^\mu}], \label{E_CAM}
\end{eqnarray}
thus leading to a multi-determinantal extension of hybrid CAM-DFT.
Note that, for  
$\alpha=0,\beta=1$ and $\alpha=1,\beta=0$, the energy expressions underlying
MP2-srDFT and RSDHf methods are recovered,
respectively.\\ 

It is essential, in order to perform practical calculations along those
lines, to provide a more explicit expression
for the complement functional $\bar{E}_{\rm Hxc}^{\rm
\mu,\alpha,\beta}[n]$ so that density functional approximations (DFAs) can be
developed. For that purpose, we rewrite Eq.~(\ref{mdEXX_alpha_beta}) as 
\begin{eqnarray}\label{eq:comp_dft_mualphabeta}
\bar{E}^{\mu,\alpha,\beta}_{\rm Hxc}[n]=E^{\rm sr,\mu}_{\rm Hxc}[n]
-\alpha\langle\Psi^\mu[n]|\hat{W}^{\rm sr,\mu}_{\rm ee}|\Psi^\mu[n]\rangle
+(1-\alpha-\beta)\langle\Psi^\mu[n]|\hat{W}^{\rm lr,\mu}_{\rm ee}|\Psi^\mu[n]\rangle
,
\end{eqnarray}
thus leading to, according to Eq.~(\ref{eq:mdHx_def}),
\begin{eqnarray}\label{eq:comp_dft_mualphabeta_2}
\bar{E}^{\mu,\alpha,\beta}_{\rm Hxc}[n]&=&(1-\alpha)E^{\rm sr,\mu}_{\rm Hxc}[n]
+\alpha E^{\rm sr,\mu}_{\rm c,md}[n]
+(1-\alpha-\beta)
\langle\Psi^\mu[n]|\hat{W}^{\rm lr,\mu}_{\rm
ee}|\Psi^\mu[n]\rangle
.
\end{eqnarray}
As mentioned previously, local
DFAs~\cite{Toulouse2004IJQC,Paziani2006PRB} have been developed for the first two
contributions on the right-hand side of
Eq.~(\ref{eq:comp_dft_mualphabeta_2}). On the other hand, the last term
needs to be further simplified.
As shown in Sec.~\ref{subsec:2bleAC}, the latter can be expressed in
terms of conventional and short-range exchange--correlation
functionals by means of a double adiabatic connection and the use of scaling
relations. 

\subsection{\sffamily \Large 
Double adiabatic connection}\label{subsec:2bleAC}

We use in this section a double adiabatic connection (AC) where the
two-electron interaction strength depends  
not only on the range-separation parameter $\mu$ (like in
range-dependent ACs~\cite{Yang:1998p441,SavRev,Teale:2010b,Teale:2011}) but also on a scaling factor $\lambda$
(like in regular linear
ACs~\cite{Teale:2009p2020,2blehybrids_Julien,2blehybrids_Fromager2011JCP,AC_2blehybrids_Yann}). This leads to the
following auxiliary equations
\begin{eqnarray}\label{nlac}
\Big(\hat{T}+\lambda\hat{W}^{\rm lr,\mu}_{\rm ee}+\hat{V}^{\mu,\lambda}
\Big)\vert\Psi^{\mu,\lambda}\rangle=\mathcal{E}^{\mu,\lambda}\vert\Psi^{\mu,\lambda}\rangle,
\end{eqnarray}
where the local potential operator 
$\hat{V}^{\mu,\lambda}=\int \ddroit{\bf r}\,v^{\mu,\lambda}({\bf r})\,\hat{n}({\bf r})$
ensures that the density constraint 
\begin{align}\label{dens_constraint}
n_{\Psi^{\mu,\lambda}}({\bf r})=n({\bf r})
\end{align}
is fulfilled for all $\lambda$ and $\mu$ values. Note that such an AC
can in principle be described accurately by using 
Legendre--Fenchel transforms in conjunction with an expansion of the local
potential in a given (finite) basis set and the computation of the partially-interacting
wavefunction at the Coupled-Cluster
level~\cite{Teale:2009p2020,Teale:2010,Teale:2010b,Teale:2011}.\\ 

Let us consider the partially long-range-interacting LL functional
\begin{align}\label{lrlambdaLL}
F^{\rm lr, \mu,\lambda}[n] = \underset{\Psi\rightarrow n}{\rm min}\langle
\Psi\vert \hat{T}+\lambda\hat{W}^{\rm lr,\mu}_{\rm ee}\vert\Psi\rangle
=
\langle\Psi^{\mu,\lambda}|\hat{T}+\lambda\hat{W}^{\rm lr,\mu}_{\rm
ee}|\Psi^{\mu,\lambda}\rangle,
\end{align}
and its KS decomposition 
\begin{eqnarray}
F^{\rm lr,\mu,\lambda}[n]
&=&T_{\rm s}[n]+E^{\rm lr,\mu,\lambda}_{\rm Hxc}[n]
.
\nonumber\\
\end{eqnarray}
Using $T_{\rm s}[n]=F^{\rm lr,\mu,0}[n]$, the partially long-range interacting Hxc energy can be expressed
as
\begin{eqnarray}
E^{\rm lr,\mu,\lambda}_{\rm Hxc}[n]
&=&\int_0^\lambda {\rm d}\nu\dfrac{{\rm d}F^{\rm lr,\mu,\nu}[n]}{{\rm d}\nu}
,
\end{eqnarray}
thus leading to, according to the Hellmann--Feynman theorem and the density
constraint in Eq.~(\ref{dens_constraint}), 
\begin{eqnarray}\label{lrHxcmulambda_int}
E^{\rm lr,\mu,\lambda}_{\rm Hxc}[n]&=&\int_0^\lambda{\rm
d}\nu\,\langle\Psi^{\mu,\nu}|\hat{W}^{\rm lr,\mu}_{\rm
ee}|\Psi^{\mu,\nu}\rangle,
\end{eqnarray}
or, equivalently,
\begin{eqnarray}\label{lrHxcmulambda_derive}
\dfrac{\partial E^{\rm lr,\mu,\lambda}_{\rm Hxc}[n]}{\partial
\lambda}&=&\langle\Psi^{\mu,\lambda}|\hat{W}^{\rm lr,\mu}_{\rm
ee}|\Psi^{\mu,\lambda}\rangle.
\end{eqnarray}
Following Toulouse {\it et al.}~\cite{IJQC06_Julien_scaling_relations}
and Yang~\cite{Yang:1998p441}, we express the
density-functional energy in Eq.~(\ref{lrHxcmulambda_int}) as  
\begin{eqnarray}
E^{\rm lr,\mu,\lambda}_{\rm Hxc}[n]&=&
\lambda \big(E_{\rm H}[n]-E^{\rm
sr,\mu}_{\rm H}[n]\big)
+
\lambda \big(E_{\rm x}[n]-E^{\rm
sr,\mu}_{\rm x}[n]\big)\nonumber\\
&&+\lambda^2\big(E_{\rm c}[n_{1/\lambda}]-E^{\rm sr,\mu/\lambda}_{\rm
c}[n_{1/\lambda}]\big)
,
\end{eqnarray}
where the scaled density $n_{1/\lambda}$ is defined
as follows
\begin{align}
n_{1/\lambda}(\mathbf{r})=(1/\lambda)^3n(\mathbf{r}/\lambda).
\end{align}
In the particular case where $\lambda=1$, $\Psi^{\mu,\lambda}$ reduces
to the long-range-interacting wavefunction $\Psi^\mu[n]$ introduced in
Eq.~(\ref{LL_parti_srdft}). We therefore obtain from
Eq.~(\ref{lrHxcmulambda_derive}),
\begin{eqnarray}
\langle\Psi^\mu[n]\vert\hat{W}^{\rm lr,\mu}_{\rm
ee}\vert\Psi^\mu[n]\rangle&=&
\left.\dfrac{\partial E^{\rm lr,\mu,\lambda}_{\rm Hxc}[n]}{\partial
\lambda}\right|_{\lambda=1}
\nonumber\\
&=&
\big(E_{\rm H}[n]-E^{\rm
sr,\mu}_{\rm H}[n]\big)
+
\big(E_{\rm x}[n]-E^{\rm
sr,\mu}_{\rm x}[n]\big)\nonumber\\
&&
+2\big(E_{\rm c}[n]-E^{\rm sr,\mu}_{\rm
c}[n]\big)
\nonumber\\
&&+
\left.\dfrac{\partial}{\partial
\lambda}
\Big(E_{\rm c}[n_{1/\lambda}]-E^{\rm sr,\mu/\lambda}_{\rm
c}[n_{1/\lambda}]\Big)
\right|_{\lambda=1},
\end{eqnarray}
thus leading to the exact expression for the complement
density-functional energy in Eq.~(\ref{eq:comp_dft_mualphabeta_2}):
\begin{eqnarray}
\bar{E}^{\mu,\alpha,\beta}_{\rm Hxc}[n]&=&
(1-\alpha-\beta)E_{\rm H}[n]+\beta E^{\rm
sr,\mu}_{\rm H}[n]
\nonumber\\
&&+
(1-\alpha-\beta)E_{\rm x}[n]+\beta E^{\rm
sr,\mu}_{\rm x}[n]
\nonumber\\
&&
+2(1-\alpha-\beta)E_{\rm c}[n]
-(1-\alpha-2\beta)E^{\rm sr,\mu}_{\rm c}[n]
+\alpha E^{\rm sr,\mu}_{\rm c,md}[n]
\nonumber\\
&&
+(1-\alpha-\beta)
\left.\dfrac{\partial}{\partial
\lambda}
\Big(E_{\rm c}[n_{1/\lambda}]-E^{\rm sr,\mu/\lambda}_{\rm
c}[n_{1/\lambda}]\Big)
\right|_{\lambda=1}.
\end{eqnarray}
As suggested by Sharkas {\it et al.}~\cite{2blehybrids_Julien} for double hybrids based on the
linear separation of the two-electron interaction, density scaling might
be neglected in practical calculations,
\begin{eqnarray}
E_{\rm c}[n_{1/\lambda}]\rightarrow E_{\rm c}[n],\hspace{0.5cm}
E^{\rm sr,\mu/\lambda}_{\rm c}[n_{1/\lambda}] \rightarrow
E^{\rm sr,\mu/\lambda}_{\rm c}[n],
\end{eqnarray}
which leads to the following approximate expression 
\begin{eqnarray}
\bar{E}^{\mu,\alpha,\beta}_{\rm Hxc}[n]&\rightarrow&
(1-\alpha-\beta)E_{\rm H}[n]+\beta E^{\rm
sr,\mu}_{\rm H}[n]
\nonumber\\
&&+
(1-\alpha-\beta)E_{\rm x}[n]+\beta E^{\rm
sr,\mu}_{\rm x}[n]
\nonumber\\
&&
+2(1-\alpha-\beta)E_{\rm c}[n]
-(1-\alpha-2\beta)E^{\rm sr,\mu}_{\rm c}[n]
+\alpha E^{\rm sr,\mu}_{\rm c,md}[n]
\nonumber\\
&&
+\mu(1-\alpha-\beta)
\frac{\partial E^{\rm sr,\mu}_{\rm c}[n]}{\partial
\mu}
.
\end{eqnarray}

\subsection{\sffamily \Large 
Coulomb-attenuating double hybrid approximation} \label{CAM-DH_approx}

In order to derive a CAM-DH scheme from the exact energy expression in Eq.~(\ref{E_CAM}), we now
introduce the modified auxiliary energy 
\begin{eqnarray}\label{newptepsilon}
{E}^{\epsilon,\mu,\alpha,\beta}
=
{E}^{\epsilon,\mu}
-\subsupi{E}{Hxc}{sr,\mu}[n_{\Psi^{\epsilon,\mu}}]
+\epsilon\dfrac{\langle\Psi^{\epsilon,\mu}|\alpha\subsupi{\hat{W}}{ee}{}+(\beta-1)\subsupi{\hat{W}}{ee}{lr,\mu}|\Psi^{\epsilon,\mu}\rangle}{\langle\Psi^{\epsilon,\mu}|\Psi^{\epsilon,\mu}\rangle}+
\bar{E}^{\mu,\alpha,\beta}_{\rm Hxc}[n_{\Psi^{\epsilon,\mu}}],
\end{eqnarray}
where ${E}^{\epsilon,\mu}$ is the original auxiliary energy underlying the
MP2-srDFT method (See Eq.~(\ref{Eepsilon_var-pt})). 
Note that both original and modified auxiliary energies are equal to the
\emph{exact} ground-state energy when $\epsilon=1$.
Following Ref.~\cite{Cornaton2013PRA} leads to the second-order expansion
\begin{eqnarray}\begin{array}{l}
\displaystyle 
{E}^{\epsilon,\mu,\alpha,\beta}
={E}^{(0)\mu,\alpha,\beta}+\epsilon{E}^{(1)\mu,\alpha,\beta}+\epsilon^2{E}^{(2)\mu,\alpha,\beta}+\mathcal{O}(\epsilon^3),
\end{array}
\end{eqnarray}
where the energy recovered through first order equals
\begin{eqnarray}
E^{(0)\mu,\alpha,\beta}+E^{(1)\mu,\alpha,\beta}=
\langle\Phi_0^\mu|\hat{T}+\alpha\subsupi{\hat{W}}{ee}{}+\beta\subsupi{\hat{W}}{ee}{lr,\mu}+\subi{\hat{V}}{ne}
|\Phi_0^\mu\rangle
+\bar{E}^{\mu,\alpha,\beta}_{\rm Hxc}[n_{\Phi_0^\mu}]
,
\end{eqnarray}
and the second-order energy correction is
\begin{eqnarray}
E^{(2)\mu,\alpha,\beta}&=&
\big(2(\alpha+\beta)-1\big)E^{(2)\rm lr,\mu}_{\rm MP}+\alpha E^{(2)\rm
lr-sr,\mu}_{\rm MP}
\nonumber\\
&&+
\int {\rm d}{\bf r}\,\Bigg(
\dfrac{\delta\bar{E}^{\mu,\alpha,\beta}_{\rm Hxc} }{\delta n({\bf r})}
-\dfrac{\delta \subsupi{E}{Hxc}{sr,\mu}}{\delta n({\bf
r})}\Bigg)[n_{\Phi_0^\mu}]\;\delta n^{(2)\mu}({\bf r}).
\end{eqnarray}
When the second-order correction to the density as well as density scaling
in the complement density-functional energy 
are neglected, a CAM-DH energy expression referred to as
d$\mu$-CAM-DH{\it lr} is obtained
\begin{eqnarray}\label{dmucam-dh-ener_exp}
E^{\rm DH{\it lr}}_{{\rm d}\mu-{\rm CAM}}
&=&
\langle\Phi_0^\mu|\hat{T}+\subi{\hat{V}}{ne}
|\Phi_0^\mu\rangle+E_{\rm H}[n_{\Phi_0^\mu}]+(\alpha+\beta) 
E^{\rm HF}_{\rm x}[\Phi_0^\mu]
-\beta
E^{\rm HF, sr, \mu}_{\rm x}[\Phi_0^\mu]
\nonumber\\
&&+
(1-\alpha-\beta)E_{\rm x}[n_{\Phi_0^\mu}]+\beta E^{\rm
sr,\mu}_{\rm x}[n_{\Phi_0^\mu}]
\nonumber\\
&&+
\big(2(\alpha+\beta)-1\big)E^{(2)\rm lr,\mu}_{\rm MP}+\alpha E^{(2)\rm
lr-sr,\mu}_{\rm MP}
\nonumber\\
&&
+2(1-\alpha-\beta)E_{\rm c}[n_{\Phi_0^\mu}]
-(1-\alpha-2\beta)E^{\rm sr,\mu}_{\rm c}[n_{\Phi_0^\mu}]
+\alpha E^{\rm sr,\mu}_{\rm c,md}[n_{\Phi_0^\mu}]
\nonumber\\
&&
+\mu(1-\alpha-\beta)
\left.\frac{\partial E^{\rm sr,\nu}_{\rm c}[n_{\Phi_0^\mu}]}{\partial
\nu}\right|_{\nu=\mu}
.
\end{eqnarray}
The corresponding expressions for the exchange and correlation energies
are
\begin{eqnarray}\label{Ex_dmu–cam-dh}
E^{\rm DH{\it lr}}_{{\rm x, d}\mu-{\rm CAM}}
&=&
(\alpha+\beta) 
E^{\rm HF}_{\rm x}[\Phi]
-\beta
E^{\rm HF, sr, \mu}_{\rm x}[\Phi]
\nonumber\\
&&+
(1-\alpha-\beta)E_{\rm x}[n]+\beta E^{\rm
sr,\mu}_{\rm x}[n],
\end{eqnarray}
and 
\begin{eqnarray}\label{Ec_dmu–cam-dh}
E^{\rm DH{\it lr}}_{{\rm c, d}\mu-{\rm CAM}}
&=&
\big(2(\alpha+\beta)-1\big)E^{(2)\rm lr,\mu}_{\rm MP}+\alpha E^{(2)\rm
lr-sr,\mu}_{\rm MP}
\nonumber\\
&&
+2(1-\alpha-\beta)E_{\rm c}[n]
-(1-\alpha-2\beta)E^{\rm sr,\mu}_{\rm c}[n]
+\alpha E^{\rm sr,\mu}_{\rm c,md}[n]
\nonumber\\
&&
+\mu(1-\alpha-\beta)
\frac{\partial E^{\rm sr,\mu}_{\rm c}[n]}{\partial
\mu}
.
\end{eqnarray}
Note that the suffix {\it lr} in d$\mu$-CAM-DH{\it lr} refers to the {\it long-range} interacting
perturbation theory this specific CAM-DH approximation relies on.
The prefix "d$\mu$" comes from the derivative with
respect to the range-separation parameter $\mu$ in the last term on the
right-hand side of Eq.~(\ref{Ec_dmu–cam-dh}). 
By neglecting this
derivative, that we call d$\mu$ correction in the following,
we obtain what we shall simply refer to as CAM-DH{\it lr} correlation energy 
\begin{eqnarray}
E^{\rm DH{\it lr}}_{\rm c, {\rm CAM}}
&=&
\big(2(\alpha+\beta)-1\big)E^{(2)\rm lr,\mu}_{\rm MP}+\alpha E^{(2)\rm
lr-sr,\mu}_{\rm MP}
\nonumber\\
&&
+2(1-\alpha-\beta)E_{\rm c}[n]
-(1-\alpha-2\beta)E^{\rm sr,\mu}_{\rm c}[n]
+\alpha E^{\rm sr,\mu}_{\rm c,md}[n]
,
\end{eqnarray}
while the CAM-DH{\it lr} exchange energy will be the same as for
d$\mu$-CAM-DH{\it lr} (see
Eq.~(\ref{Ex_dmu–cam-dh})).

Interestingly, CAM-DH{\it lr} reduces to
MP2-srDFT and
RSDHf when 
$\alpha=0,\beta=1$ and $\alpha=1,\beta=0$, respectively. Using
the linearity in $\alpha$ and $\beta$ of the CAM-DH{\it lr} correlation energy
leads to the
compact expression
\begin{eqnarray}
E^{\rm DH{\it lr}}_{\rm c, {\rm CAM}}
&=&
\alpha E_{\rm c,RSDHf}
+\beta E^{\rm srDFT}_{\rm c, MP2}
+(1-\alpha-\beta)\Big(2E_{\rm c}[n]-E^{\rm sr,\mu}_{\rm
c}[n]-E^{(2)\rm lr,\mu}_{\rm MP}\Big).
\label{E_CAM-DH}
\end{eqnarray}
Note finally that, in the particular case where $\alpha+\beta=1$,
CAM-DH{\it lr}
reduces to the {\it two-parameter} RSDHf (2RSDHf) scheme introduced in
the Appendix A of Ref.~\cite{Cornaton2013PRA} The second parameter
(referred to as $\lambda$ in Ref.~\cite{Cornaton2013PRA})  
is here equal to $\alpha$.


\section{\sffamily \Large COMPUTATIONAL DETAILS}\label{sec:comp_details}





The (d$\mu$-)CAM-DH{\it lr} exchange and correlation energies in
Eqs.~(\ref{Ex_dmu–cam-dh}), (\ref{Ec_dmu–cam-dh}) and (\ref{E_CAM-DH})
have been computed with a development version of the DALTON program
package~\cite{dalton_paper}. Spin-unpolarized LDA-type functionals~\cite{Toulouse2004IJQC,Paziani2006PRB} 
have been used for modeling complement density-functional
energy contributions. The corresponding double hybrid approximations
will therefore be referred to with the suffix LDA:
(d$\mu$-)CAM-DH{\it lr}-LDA, MP2-srLDA, RSDHf-LDA and 2RSDHf-LDA. 
The range-separation parameter was set to the prescribed value $\mu=0.4
{a_0}^{-1}$ (see Ref.~\cite{Cornaton2013PRA}).
Augmented
correlation-consistent polarized quadruple-$\zeta$ basis sets
("aug-cc-pVQZ") of Dunning and
co-workers~\cite{Dunning1989JCP,Kendall1992JCP,Woon1993JCP,Woon1994JCP,Koput2002JPCA,Wilson1999JCP}
have been used. 
Interaction energies have been computed for the first three noble-gas
homonuclear dimers ($\rm He_2$, $\rm Ne_2$ and $\rm Ar_2$) as well as
for the first two homonuclear alkaline-earth-metal dimers ($\rm Be_2$
and $\rm Mg_2$). Since both CAM and range-separated double hybrid
schemes considered in this work rely
on a long-range-interacting only perturbation theory, BSSE is expected
to  
be small~\cite{Cornaton2013PRA}. Consequently, no BSSE correction was
made.
In addition, 
total energies have been computed around the equilibrium distance for $\rm H_2$, $\rm Li_2$, $\rm C_2$,
$\rm N_2$ and $\rm F_2$.

\section{\sffamily \Large RESULTS AND DISCUSSION}\label{sec:discussion}




\subsection{Results for He$_2$, Ne$_2$ and Ar$_2$}\label{subsec:Rg2}

As mentioned in Sec.~\ref{CAM-DH_approx}, in the particular case where
$\alpha+\beta=1$, CAM-DH{\it lr}-LDA reduces to the 2RSDHf-LDA
scheme of Ref.~\cite{Cornaton2013PRA}, where the second parameter equals
$\lambda=\alpha=1-\beta$. The 2RSDHf-LDA energy is in fact the weighted
average value of RSDHf-LDA and MP2-srLDA energies with weights $\lambda$
and $(1-\lambda)$, respectively.
Therefore, the long-range correlation energy is entirely described
within MP2. On the other hand, the coupling between long- and
short-range correlations is decomposed into MP2 and
density-functional contributions. Obviously, when applied to weakly
interacting systems, 2RSDHf-LDA can only improve MP2-srLDA and RSDHf-LDA
interaction energies around
the equilibrium distance where the long-range--short-range MP2 coupling term
can be significant~\cite{Cornaton2013PRA}. 
For $0<\lambda<1$, the 2RSDHf-LDA curve will be located between the
MP2-srLDA and RSDHf-LDA ones. The latter are shown in Fig.~\ref{compar}.
Consequently, the agreement with experiment
strongly depends on the performance of the MP2-srLDA and RSDHf-LDA methods. For
$\rm He_2$ and $\rm Ne_2$, they both underestimate the interaction
energy. In this particular case, 2RSDHf-LDA will bring no improvement relative
to MP2-srLDA and RSDHf-LDA. In $\rm Ar_2$, RSDHf-LDA overbinds while MP2-srLDA
slightly underbinds. It is then possible to find a $\lambda$ value for
which the 2RSDHf-LDA interaction energy equals the
experimental one. However, the equilibrium bond distance will then be
overestimated and the interaction energy at long
distance will remain overestimated (in absolute value).
According to Ref.~\cite{PRA10_Julien_rpa-srDFT}, substituting a
long-range {\it Random-Phase Approximation} (RPA) for
the long-range MP2 treatment may improve the potential curve at large distance. A
more pragmatic alternative, that we investigate further in the
following, consists in treating only a fraction of the long-range
correlation energy within MP2.

We will therefore relax the condition $\alpha+\beta=1$, in analogy with 
the hybrid CAM-B3LYP functional~\cite{Yanai2004CPL} where the
Coulomb attenuation is used for the exchange energy only. Even though we
are using the former for both exchange and correlation energies, it is
interesting for analysis purposes to use the same parameters as in
CAM-B3LYP ($\alpha=0.19$ and $\beta=0.46$). 
Unlike the value $\mu=0.4$ that is based on the analysis of long-range
correlation effects~\cite{Cornaton2013PRA}, these two values have been optimized
empirically in a completely different context. In this respect, the
corresponding CAM-DH{\it lr}-LDA approach is semi-empirical.
In contrast to
2RSDHf-LDA, the CAM-DH{\it lr}-LDA energy includes then a third term, in
addition to the total RSDHf-LDA
and MP2-srLDA energies. Indeed, according to Eqs.~(\ref{Ex_dmu–cam-dh})
and (\ref{E_CAM-DH}), the former can be
rewritten as
\begin{eqnarray}\label{eq:CAM-DH_ener_3terms}
E^{\rm DH{\it lr}-LDA}_{\rm CAM}&=&(1-\alpha-\beta)\big(
E^0_{\rm LDA}+E^{\rm LDA}_{\rm c}[n]-E_{\rm c}^{\rm srLDA,\mu}[n]-E_{\rm MP}^{\rm (2)lr,\mu}
\big)
\nonumber\\
&& +\alpha E^{\rm LDA}_{\rm RSDHf}+\beta E^{\rm srLDA}_{\rm MP2},
\end{eqnarray}
where $n$ denotes here the HF-srLDA density and $E^0_{\rm LDA}$ is the
conventional KS-LDA total energy computed with the HF-srLDA determinant.  
Let us stress that, in this case, only a fraction $2(\alpha+\beta)-1=0.3$ of
the long-range correlation energy is described by MP2. 

Each contribution to the first term on the right-hand side of
Eq.~(\ref{eq:CAM-DH_ener_3terms}) as well as the total CAM-DH{\it
lr}-LDA
interaction energy have been computed for the three dimers.
Results are shown in Fig.~\ref{compar}. As
expected~\cite{vanMourick2002JCP}, the total LDA energy contribution
$E^0_{\rm LDA}$ is always too attractive.   
For $\rm He_2$ and $\rm Ne_2$, the remaining contributions do not
compensate this large error, which explains why CAM-DH{\it lr}-LDA strongly
overbinds. On the other hand, for $\rm Ar_2$, the overbinding induced by
the total LDA energy contribution is less pronounced so that the
significant long-range MP2 term compensates the error and leads to a
CAM-DH{\it lr}-LDA curve that is significantly less attractive relative to MP2-srLDA
and RSDHf-LDA. Note also that the potential curves are less accurate at
long distance relative to MP2-srLDA, RSDHf-LDA and
2RSDHf-LDA. 

These observations suggest that the parameters $\alpha=0.19$ and
$\beta=0.46$ optimized for the CAM-B3LYP functional are not optimal in
this context. As shown in Fig.~\ref{ab}, it is possible to tune
$\alpha$ and $\beta$ in order to obtain more accurate potential energy
curves.
Choosing $\alpha+\beta$ slightly smaller than 1 (between 0.8 and 0.9)
seems to give the best results. Note that, in that case, CAM-DH{\it
lr}-LDA performs
much better than MP2-srLDA and RSDHf-LDA for $\rm Ar_2$ at long distance. 
As discussed further in
Sec.~\ref{CAM-DH_alternative}, it would also be interesting to test
another formulation of CAM-DH where the perturbation expansion of the
wavefunction is based on a partially long-range interacting system. In
such an approach, the CAM-DH energy is not linear in $\alpha$ and
$\beta$ anymore.    


Let us finally discuss the performance of the d$\mu$-CAM-DH{\it lr}-LDA method that
simply consists in adding to the CAM-DH{\it lr}-LDA energy the fraction $\mu(1-\alpha-\beta)$ of the
first-order derivative $\partial E_{\rm c}^{\rm srLDA,\mu}[n]/\partial \mu$ at 
$\mu=0.4 {a_0}^{-1}$.
Fig.~\ref{Ecsr_mu} shows the variation with $\mu$ of the srLDA
correlation interaction energy computed with the HF-srLDA
($\mu=0.4$) density for
different bond distances in $\rm He_2$. That contribution is clearly linear in
the vicinity of $0.4 {a_0}^{-1}$. It is therefore relevant to
approximate the first-order derivative as follows
\begin{eqnarray}\label{Ealphabeta}
\left.\dfrac{\partial E_{\rm c}^{\rm srLDA,\mu}[n]}{\partial
\mu}\right|_{\mu=0.4 {a_0}^{-1}}&\approx&\dfrac{E_{\rm c}^{\rm
srLDA,\mu=0.405 {a_0}^{-1}}[n]-E_{\rm c}^{\rm srLDA,\mu=0.395
{a_0}^{-1}}[n]}{0.01}.
\end{eqnarray}
Fig 3. in Ref.~\cite{Cornaton2013PRA} suggests that this approximation is also relevant for
the other dimers. 
Results obtained for the semi-empirical $\alpha=0.19$ and $\beta=0.46$
parameters are shown in Fig.~\ref{dmu}.
As
expected from Fig.~\ref{Ecsr_mu} (where the slope of the srLDA
correlation interaction energy is always positive), d$\mu$-CAM-DH{\it
lr}-LDA binds
less than CAM-DH{\it lr}-LDA. The difference is quite significant which is  
an improvement for $\rm He_2$ and $\rm Ne_2$ but not for $\rm Ar_2$. 

\subsection{Results for Be$_2$ and Mg$_2$}

Interaction energies have been computed for the first two
alkaline-earth-metal dimers. The
beryllium dimer is difficult to describe with DFT-based methods because (i)
dispersion forces bind the two atoms and (ii) the
latter exhibit significant multiconfigurational effects due to the
low-lying 2p orbitals~\cite{Stark1996CPL}. A multireference
extension~\cite{nevpt2srdft} of CAM-DH{\it lr} would actually be more appropriate in this
context. This is left for future work.\\

Here we discuss the interaction energy curves obtained at the CAM-DH{\it
lr}-LDA
level. 
Results are shown in Fig.~\ref{compar-weak} and comparison is made with MP2-srLDA
and RSDHf-LDA. We first note that, in contrast to the rare-gas
dimers, both Be$_2$ and Mg$_2$ have equilibrium interaction energies
that are larger in absolute value at the MP2-srLDA level, relative to
RSDHf-LDA. Both methods underbind while CAM-DH{\it lr}-LDA, using the 
semi-empirical $\alpha=0.19$ and
$\beta=0.46$ parameters, overbinds. In the latter
case, the energy contribution that is recovered when $\alpha=0,\beta=0$ (first term
on the right-hand side of Eq.~(\ref{eq:CAM-DH_ener_3terms})) is too attractive and the
scaling factor $1-0.19-0.46=0.35$ is large enough to induce overbinding.   
Like in the rare-gas dimers, using $\alpha=0.2,\beta=0.7$ or
$\alpha=0.6,\beta=0.3$ provides reasonable equilibrium interaction
energies. On the other hand, no improvement is observed at long
distance.\\

Finally, the d$\mu$ correction has been computed when $\alpha=0.19$ and
$\beta=0.46$. As observed for the rare-gas dimers, it reduces the
equilibrium interaction energy which is an improvement for Be$_2$ but not for
Mg$_2$. The performance of the d$\mu$-CAM-DH{\it lr}-LDA scheme obviously depends on the choice of $\alpha$ and
$\beta$. 
Density
scaling in the correlation functionals is also expected to be
important~\cite{2blehybrids_Julien,metaGGA-2blehybrids_Julien}. 
Enlarging the test set and fitting all parameters on experimental data would be necessary at this
point. 

\subsection{Results for H$_2$, Li$_2$, C$_2$, N$_2$ and F$_2$}

We now consider diatomics with single $\sigma$ bond ($\rm H_2$, $\rm Li_2$ and $\rm F_2$), triple $\sigma+\pi+\pi$ bond ($\rm N_2$) and even the unusual double $\pi+\pi$ bond ($\rm C_2$).
Note that, in order to
describe the dissociation regime, a multiconfiguration hybrid CAM-DFT
approach should be developed, in the spirit of multiconfiguration DFT
based on the linear~\cite{Sharkas_JCP} and
range~\cite{PedersenThesis,dft-Fromager-JCP2007a,JCPunivmu2}
separations of the two-electron repulsion.
We focus here on the total energies around the equilibrium distances.
Potential energy curves are shown in Fig.~\ref{compar-covalent}.
While MP2-srLDA understimates the total energies in absolute value,
RSDHf-LDA energies are too low. CAM-DH{\it lr}-LDA is slightly more accurate than
MP2-srLDA  when the 
semi-empirical $\alpha=0.19$ and
$\beta=0.46$ parameters are used. Note that the d$\mu$
correction to the total energy is positive, thus leading to a
d$\mu$-CAM-DH{\it lr}-LDA energy that is
higher than the CAM-DH{\it lr}-LDA one and therefore less accurate for these systems.   
Let us mention that Toulouse {\it et al.} already observed in the helium
atom that the srLDA correlation energy has a positive slope at $\mu=0.4$
(see Fig.~6 in Ref.~\cite{Toulouse2004PRA}). Note that CAM-DH{\it
lr}-LDA total
energies obtained with the parameters $\alpha=0.6,\beta=0.3$ (see
Sec.~\ref{subsec:Rg2})
are in relatively good agreement with the accurate values. 




\section{\sffamily \Large PERSPECTIVE: ALTERNATIVE FORMULATION OF
THE CAM-DH APPROXIMATION} \label{CAM-DH_alternative}

All the double hybrid energy expressions derived and tested previously rely on the perturbation
expansion of a fully long-range interacting wavefunction. This choice
was motivated by the fact that the long-range interaction based on the
error function has no
singularity at $r_{12}=0$. Consequently, the BSSE is significantly reduced and the
convergence with respect to the atomic basis set is faster relative to
regular MP2 (see Ref.~\cite{Cornaton2013PRA} and the references therein). Such features would
actually be preserved if a partially long-range interacting wavefunction
was used instead. This choice seems in fact more sound since,
within a CAM-DH scheme, we aim at describing only a fraction
$(\alpha+\beta)$ of the
long-range interaction within MP2.  

In order to derive such an alternative CAM-DH, we first consider the following decomposition of the
universal LL functional,
\begin{eqnarray}\label{eq:HKfundecomp}
F[n]&=&F^{\rm lr,+\infty,1}[n]
\nonumber\\
&=& F^{\rm lr,+\infty,\lambda}[n]
+(1-\lambda)\big(E_{\rm H}[n]+E_{\rm x}[n]\big)+E_{\rm
c}[n]-\lambda^2E_{\rm c}[n_{1/\lambda}].
\end{eqnarray}
The latter relies on the linear separation of the two-electron
repulsion~\cite{AC_2blehybrids_Yann}. We
then separate the partially-interacting LL functional into long- and
short-range parts
\begin{eqnarray}
F^{\rm lr,+\infty,\lambda}[n]
=
F^{\rm lr,\mu,\lambda}[n]+
E_{\rm Hxc}^{\rm
sr,\mu,\lambda}[n],
\end{eqnarray}
where, according to
Refs.~\cite{IJQC06_Julien_scaling_relations,Yang:1998p441}, 
\begin{eqnarray}\label{srfunlambdamu}
E_{\rm Hxc}^{\rm
sr,\mu,\lambda}[n]=
\lambda\big(E^{\rm sr,\mu}_{\rm H}[n]+E^{\rm sr,\mu}_{\rm x}[n]\big)
+\lambda^2E^{\rm sr,\mu/\lambda}_{\rm c}[n_{1/\lambda}]
.
\end{eqnarray}
We finally obtain from Eq.~(\ref{eq:HKfundecomp}) the following
decomposition for the LL functional
\begin{eqnarray}
F[n]=F^{\rm lr,\mu,\lambda}[n]+\bar{E}_{\rm Hxc}^{\mu,\lambda}[n]
,
\end{eqnarray}
where the complement density-functional energy equals
\begin{eqnarray}\label{eq:comp_dft_two_param}
\bar{E}_{\rm Hxc}^{\mu,\lambda}[n]
&=&
\lambda\big(E^{\rm sr,\mu}_{\rm H}[n]+E^{\rm sr,\mu}_{\rm x}[n]\big)
+(1-\lambda)\big(E_{\rm H}[n]
+E_{\rm x}[n]\big)
\nonumber\\
&&
+\lambda^2E^{\rm sr,\mu/\lambda}_{\rm c}[n_{1/\lambda}]
+E_{\rm
c}[n]-\lambda^2E_{\rm c}[n_{1/\lambda}].
\end{eqnarray}
According to the variational principle in Eq.~(\ref{VarprincipleHK}), the {\it exact}
ground-state energy can then be rewritten as
\begin{eqnarray}\label{eq:2srDFT_ener}
E&=&\min_\Psi\left\{
\langle\Psi|\hat{T}+\hat{V}_{\rm
ne}+\lambda\hat{W}^{\rm lr,\mu}_{\rm
ee}|\Psi\rangle+\bar{E}^{\mu,\lambda}_{\rm Hxc}[n_{\Psi}]
\right\}
\nonumber\\
&=&
\langle\tilde{\Psi}^{\mu,\lambda}|\hat{T}+\hat{V}_{\rm
ne}+\lambda\hat{W}^{\rm lr,\mu}_{\rm
ee}|\tilde{\Psi}^{\mu,\lambda}\rangle+\bar{E}^{\mu,\lambda}_{\rm
Hxc}[n_{\tilde{\Psi}^{\mu,\lambda}}],
\end{eqnarray}
where the minimizing wavefunction $\tilde{\Psi}^{\mu,\lambda}$ is the
ground state of the partially long-range interacting system whose
density equals the {\it exact} ground-state density of the physical system. 
A new class of range-separated double hybrids can then be formulated
when solving Eq.~(\ref{eq:2srDFT_ener}) with many-body perturbation
theory techniques.

By analogy with the HF-srDFT approximation, we obtain a {\it
two-parameter range-separated hybrid} (2RSH) determinant, 
\begin{align}
\Phi^{\mu,\lambda}_0=\!\argmin\limits_{\Phi} \!\left\{\!\langle \Phi \vert
\hat{T}+\lambda\hat{W}^{\rm lr,\mu}_{\rm ee}+\hat{V}_{\rm
ne}\vert\Phi\rangle+\bar{E}^{\mu,\lambda}_{\rm Hxc}[n_{\Phi}] \right\},
\end{align}
when restricting the minimization in Eq.~(\ref{eq:2srDFT_ener}) to
single determinants $\Phi$. Using a MP-type perturbation
theory~\cite{Angyan2005PRA,Fromager2008PRA,pra_MBPTn-srdft_janos} in this context leads
to the following perturbation expansion for the energy
\begin{align}
E
=
\langle\Phi^{\mu,\lambda}_0|\hat{T}+\hat{V}_{\rm
ne}+\lambda\hat{W}^{\rm lr,\mu}_{\rm
ee}|\Phi^{\mu,\lambda}_0\rangle+\bar{E}^{\mu,\lambda}_{\rm
Hxc}[n_{\Phi^{\mu,\lambda}_0}]
+\lambda^2E^{(2)\rm lr,\mu}_{\rm MP}+\ldots
\end{align}
If, instead, we split the complement density functional into 
wavefunction and density-functional terms as follows
\begin{eqnarray}\label{newsplitcompfun}
\bar{E}^{\mu,\lambda}_{\rm
Hxc}[n]=\alpha
\langle\Psi^{\mu,\lambda}[n]\vert\hat{W}^{\rm sr,\mu}_{\rm
ee}\vert\Psi^{\mu,\lambda}[n]\rangle
+\bar{E}^{\mu,\lambda,\alpha}_{\rm Hxc}[n],
\end{eqnarray}
where, according to the Appendix,
\begin{eqnarray}\label{complDFTfun_mulambdaalpha}
\bar{E}^{\mu,\lambda,\alpha}_{\rm Hxc}[n]
&=& 
(\lambda-\alpha)\big(E^{\rm sr,\mu}_{\rm H}[n]+E^{\rm sr,\mu}_{\rm x}[n]\big)
+(1-\lambda)\big(E_{\rm H}[n]
+E_{\rm x}[n]\big)
\nonumber\\
&&
+\lambda(\lambda-\alpha)E^{\rm sr,\mu/\lambda}_{\rm c}[n_{1/\lambda}]
+E_{\rm
c}[n]-\lambda^2E_{\rm c}[n_{1/\lambda}]
\nonumber\\
&&
+\alpha\lambda E^{\rm sr,\mu/\lambda}_{\rm c, md}[n_{1/\lambda}],
\end{eqnarray}
the {\it exact} ground-state energy can be rewritten, according to
Eq.~(\ref{eq:2srDFT_ener}), as 
\begin{eqnarray}
E&=&
\langle\tilde{\Psi}^{\mu,\lambda}|\hat{T}+\hat{V}_{\rm
ne}+\lambda\hat{W}^{\rm lr,\mu}_{\rm
ee}+\alpha\hat{W}^{\rm sr,\mu}_{\rm
ee}|\tilde{\Psi}^{\mu,\lambda}\rangle+\bar{E}^{\mu,\lambda,\alpha}_{\rm
Hxc}[n_{\tilde{\Psi}^{\mu,\lambda}}]
,
\end{eqnarray}
since
$\Psi^{\mu,\lambda}[n_{\tilde{\Psi}^{\mu,\lambda}}]=\tilde{\Psi}^{\mu,\lambda}$.\\

A MP2-type perturbation theory similar to the one derived in
Sec.~\ref{CAM-DH_approx}
can then be formulated, thus leading to the following perturbation
expansion for the energy through second order 
\begin{eqnarray}
E&=&
\langle\Phi^{\mu,\lambda}_0|\hat{T}+\hat{V}_{\rm
ne}+\lambda\hat{W}^{\rm lr,\mu}_{\rm
ee}+\alpha\hat{W}^{\rm sr,\mu}_{\rm
ee}|\Phi^{\mu,\lambda}_0\rangle+\bar{E}^{\mu,\lambda,\alpha}_{\rm
Hxc}[n_{\Phi^{\mu,\lambda}_0}]
\nonumber\\
&&+\lambda^2E^{(2)\rm lr,\mu}_{\rm MP}+\alpha\lambda E^{(2)\rm
lr-sr,\mu}_{\rm MP}+\ldots,
\end{eqnarray}
where the second-order correction to the density has been neglected.
Note that the long-range and long-range--short-range MP2 correlation
energies are obtained from the 2RSH orbitals and
orbital energies. Therefore, they depend implicitly on $\lambda$. 

Thus, we obtain an alternative CAM-DH 
approximation which can be compared with the d$\mu$-CAM-DH{\it lr} energy expression in
Eq.~(\ref{dmucam-dh-ener_exp})
when choosing $\lambda=\alpha+\beta$. 
Interestingly, the fractions of long-range and
long-range--short-range MP2
correlation energies are now quadratic in $\alpha$ and $\beta$. They are equal to $(\alpha+\beta)^2$ and
$\alpha(\alpha+\beta)$, respectively. The implementation and calibration
of this approach that we could refer to as CAM-DH{\it plr}, where $plr$
refers to the {\it partially long-range} interacting perturbation theory
it relies on, is left for future work.

\section{\sffamily \Large CONCLUSIONS}\label{sec:conclusions}

The rigorous formulation of Coulomb-attenuating double-hybrid methods (CAM-DH)
has been investigated. In order to
preserve the advantages of existing range-separated double hybrids
(relatively fast convergence with respect to the basis set, small BSSE),
we opted for a CAM-DH that relies on the perturbation expansion of a 
long-range interacting wavefunction, in the spirit of
Ref.~\cite{Cornaton2013PRA} The method has been tested within the local
density approximation on a small test
set consisting of rare-gas and alkaline-earth-metal dimers as well as
diatomics with single, double and triple bonds. In this
context,
the semi-empirical $\alpha=0.19$ and
$\beta=0.46$ CAM parameters, that were optimized for the hybrid
CAM-B3LYP functional, do not provide accurate interaction and total
energies. Better results are obtained when $\alpha+\beta$ is closer (but
not equal) to 1, at least within the formulation we opted for.
Calibration studies should be performed on a larger test set. 
The benzene dimer and charge-transfer
complexes ({\it e.g.} $\rm HCN\cdots NH_3$~\cite{Steinmann2012JCTC})
would be good candidates. Work is in progress in this direction.
Note that density scaling in the complement correlation functional has
not been taken into account in this work, though the effect is expected
to be important~\cite{2blehybrids_Julien}. This as well as the
construction of CAM-DH using semi-local complement
functionals~\cite{2blehybrids_Julien,metaGGA-2blehybrids_Julien} should obviously be
investigated further in the future.
All parameters could be fitted on experimental data but,
in the
light of Ref.~\cite{AC_2blehybrids_Yann}, it would
also be interesting for rationalizing the fitted parameters to derive and compute the approximate double adiabatic connection
underlying CAM-DH and compare with accurate {\it ab initio}
calculations. 
An alternative formulation of CAM-DH has
finally been discussed as a perspective. It relies on the perturbation
expansion of a partially long-range interacting wavefunction. In
contrast to the CAM-DH tested in this work, the correlation energy is
not linear in $\alpha$ and $\beta$ anymore. The implementation and
calibration of such an approach is left for future work. 

\subsection*{\sffamily \large ACKNOWLEDGMENTS}

The authors would like to thank Yann Schaerer and Julien Toulouse for fruitful 
discussions. EF thanks ANR (DYQUMA project) and LabEx "Chimie des
Syst\`{e}mes Complexes" for funding. 



\renewcommand{\theequation}{A\arabic{equation}}
    \setcounter{equation}{0}  

\section*{\sffamily \large APPENDIX: scaling relation for the
multi-determinantal short-range exact exchange energy}

By analogy with Eq.~(\ref{eq:mdHx_def}), we consider the following decomposition of the
partial short-range density-functional energy
\begin{eqnarray}\label{mdsrexxlambda}
E_{\rm Hxc}^{\rm
sr,\mu,\lambda}[n]
=\lambda
\langle\Psi^{\mu,\lambda}[n]\vert\hat{W}^{\rm sr,\mu}_{\rm
ee}\vert\Psi^{\mu,\lambda}[n]\rangle
+E^{\rm sr,\mu,\lambda}_{\rm c,
md}[n].
\end{eqnarray}
Since, according to Ref.~\cite{IJQC06_Julien_scaling_relations},
\begin{eqnarray}
\Psi^{\mu\gamma}[n_\gamma]=\Psi_\gamma^{\mu,1/\gamma}[n],
\end{eqnarray}
where for any $N$-electron wavefunction $\Psi$,
\begin{eqnarray}
\Psi_\gamma({\bf{r}}_1,\ldots,{\bf{r}}_N)=\gamma^{3N/2}\Psi(\gamma{\bf{r}}_1,\ldots,\gamma{\bf{r}}_N),
\end{eqnarray}
we obtain the following scaling relation
\begin{eqnarray}
\langle\Psi^{\mu\gamma}[n_\gamma]\vert\hat{W}^{\rm sr,\mu\gamma}_{\rm
ee}\vert\Psi^{\mu\gamma}[n_\gamma]\rangle
=
\gamma
\langle\Psi^{\mu,1/\gamma}[n]\vert\hat{W}^{\rm sr,\mu}_{\rm
ee}\vert\Psi^{\mu,1/\gamma}[n]\rangle,
\end{eqnarray}
which, according to Eq.~(\ref{srfunlambdamu}) as well as Eqs.~(17) and
(18) in Ref.~\cite{IJQC06_Julien_scaling_relations}, leads to
\begin{eqnarray}
\gamma^2 E^{\rm sr,\mu,1/\gamma}_{\rm c,
md}[n]&=&\gamma^2\Big(
E_{\rm Hxc}^{\rm
sr,\mu,1/\gamma}[n]-
\dfrac{1}{\gamma}
\langle\Psi^{\mu,1/\gamma}[n]\vert\hat{W}^{\rm sr,\mu}_{\rm
ee}\vert\Psi^{\mu,1/\gamma}[n]\rangle
\Big)
\nonumber\\
&=&E^{\rm sr,\mu\gamma}_{\rm Hxc}[n_\gamma]-
\langle\Psi^{\mu\gamma}[n_\gamma]\vert\hat{W}^{\rm sr,\mu\gamma}_{\rm
ee}\vert\Psi^{\mu\gamma}[n_\gamma]\rangle
\nonumber\\
&=&E^{\rm sr,\mu\gamma}_{\rm c,md}[n_\gamma].
\end{eqnarray}
In the particular case where $\gamma=1/\lambda$, we obtain from Eqs.~(\ref{srfunlambdamu}) and~(\ref{mdsrexxlambda}) 
\begin{eqnarray}
\lambda
\langle\Psi^{\mu,\lambda}[n]\vert\hat{W}^{\rm sr,\mu}_{\rm
ee}\vert\Psi^{\mu,\lambda}[n]\rangle
&=&
E_{\rm Hxc}^{\rm
sr,\mu,\lambda}[n]
-E^{\rm sr,\mu,\lambda}_{\rm c,
md}[n]
\nonumber\\
&=&
\lambda\big(E^{\rm sr,\mu}_{\rm H}[n]+E^{\rm sr,\mu}_{\rm x}[n]\big)
+\lambda^2E^{\rm sr,\mu/\lambda}_{\rm c}[n_{1/\lambda}]
\nonumber\\
&&
-\lambda^2E^{\rm sr,\mu/\lambda}_{\rm c,md}[n_{1/\lambda}]
,
\end{eqnarray}
or, equivalently,
\begin{eqnarray}\label{Hxmd_as_fun}
\langle\Psi^{\mu,\lambda}[n]\vert\hat{W}^{\rm sr,\mu}_{\rm
ee}\vert\Psi^{\mu,\lambda}[n]\rangle
=
E^{\rm sr,\mu}_{\rm H}[n]+E^{\rm sr,\mu}_{\rm x}[n]
+\lambda E^{\rm sr,\mu/\lambda}_{\rm c}[n_{1/\lambda}]
-\lambda E^{\rm sr,\mu/\lambda}_{\rm c,md}[n_{1/\lambda}]
.
\end{eqnarray}
Combining Eq.~(\ref{Hxmd_as_fun}) with
Eqs.~(\ref{eq:comp_dft_two_param}) and~(\ref{newsplitcompfun}) leads
to Eq.~(\ref{complDFTfun_mulambdaalpha}).

\clearpage





\providecommand{\Aa}[0]{Aa}



\clearpage

\begin{figure}
\caption{\label{compar} Interaction energy curves obtained at the
CAM-DH{\it lr}-LDA
level for $\rm He_2$ (a), $\rm Ne_2$ (b) and $\rm Ar_2$ (c) with
$\mu=0.4$, $\alpha=0.19$ and $\beta=0.46$. Comparison is made with
RSDHf-LDA and
MP2-srLDA results. Various contributions to the CAM-DH{\it lr}-LDA interaction
energy are also plotted. See text for further details. CAM-DH and RSDHf
are here shorthand notations
for CAM-DH{\it lr}-LDA and RSDHf-LDA, respectively. 
The experimental curves are from Ref.~\cite{Tang2003JCP}}
\end{figure}

\begin{figure}
\caption{\label{ab} CAM-DH{\it lr}-LDA interaction energy curves obtained for $\rm
He_2$ (a), $\rm Ne_2$ (b) and $\rm Ar_2$ (c) 
with $\alpha=0.2,\beta=0.7$ and $\alpha=0.6,\beta=0.3$. 
The $\mu$ parameter was set to $0.4a_0^{-1}$.
Comparison is made with
experiment~\cite{Tang2003JCP}. RSDHf is here shorthand for RSDHf-LDA.}
\end{figure}

\begin{figure}
\caption{\label{Ecsr_mu} Short-range LDA correlation energy contribution
to the interaction energy of $\rm He_2$ obtained for three bond
distances when varying $\mu$
with fixed HF-srLDA ($\mu=0.4$) densities.
See text for further details.}
\end{figure}

\begin{figure}
\caption{\label{dmu} Interaction energy curves obtained at the
d$\mu$-CAM-DH{\it lr}-LDA level for $\rm He_2$ (a), $\rm Ne_2$ (b) and $\rm Ar_2$
(c) with $\mu=0.4$, $\alpha=0.19$ and $\beta=0.46$. Comparison is made
with CAM-DH{\it lr}-LDA (using the same parameters) and experiment~\cite{Tang2003JCP}.
d$\mu$-CAM-DH and CAM-DH are here shorthand notations for
d$\mu$-CAM-DH{\it lr}-LDA and CAM-DH{\it lr}-LDA, respectively.
}
\end{figure}

\begin{figure}
\caption{\label{compar-weak} Interaction energy curves obtained at the
CAM-DH{\it lr}-LDA and d$\mu$-CAM-DH{\it lr}-LDA
levels for $\rm Be_2$ (a) and $\rm Mg_2$ (b) 
with $\mu=0.4$.
Comparison is made with RSDHf-LDA (simply denoted RSDHf here) and
MP2-srLDA results. Various contributions to the CAM-DH{\it lr}-LDA interaction
energy are also plotted.
d$\mu$-CAM-DH and CAM-DH are here shorthand notations for
d$\mu$-CAM-DH{\it lr}-LDA and CAM-DH{\it lr}-LDA, respectively.
The accurate curves are from Refs.~\cite{Roeggen2005IJQC,Balfour1970CJP} See text for further details.}
\end{figure}

\begin{figure}
\caption{\label{compar-covalent} Total energy curves obtained at the
CAM-DH{\it lr}-LDA and d$\mu$-CAM-DH{\it lr}-LDA
levels for $\rm H_2$ (a), $\rm Li_2$ (b), $\rm C_2$ (c), $\rm N_2$ (d) and $\rm F_2$ (e) 
with $\mu=0.4$.
d$\mu$-CAM-DH and CAM-DH are here shorthand notations for
d$\mu$-CAM-DH{\it lr}-LDA and CAM-DH{\it lr}-LDA, respectively.
Comparison is made with RSDHf-LDA (simply denoted RSDHf here) and
MP2-srLDA results. 
The accurate curves are from Ref.~\cite{LieJCP1974} See text for further details.}
\end{figure}



\clearpage

%

\begin{center}
\includegraphics[width=0.85\columnwidth,keepaspectratio=true]{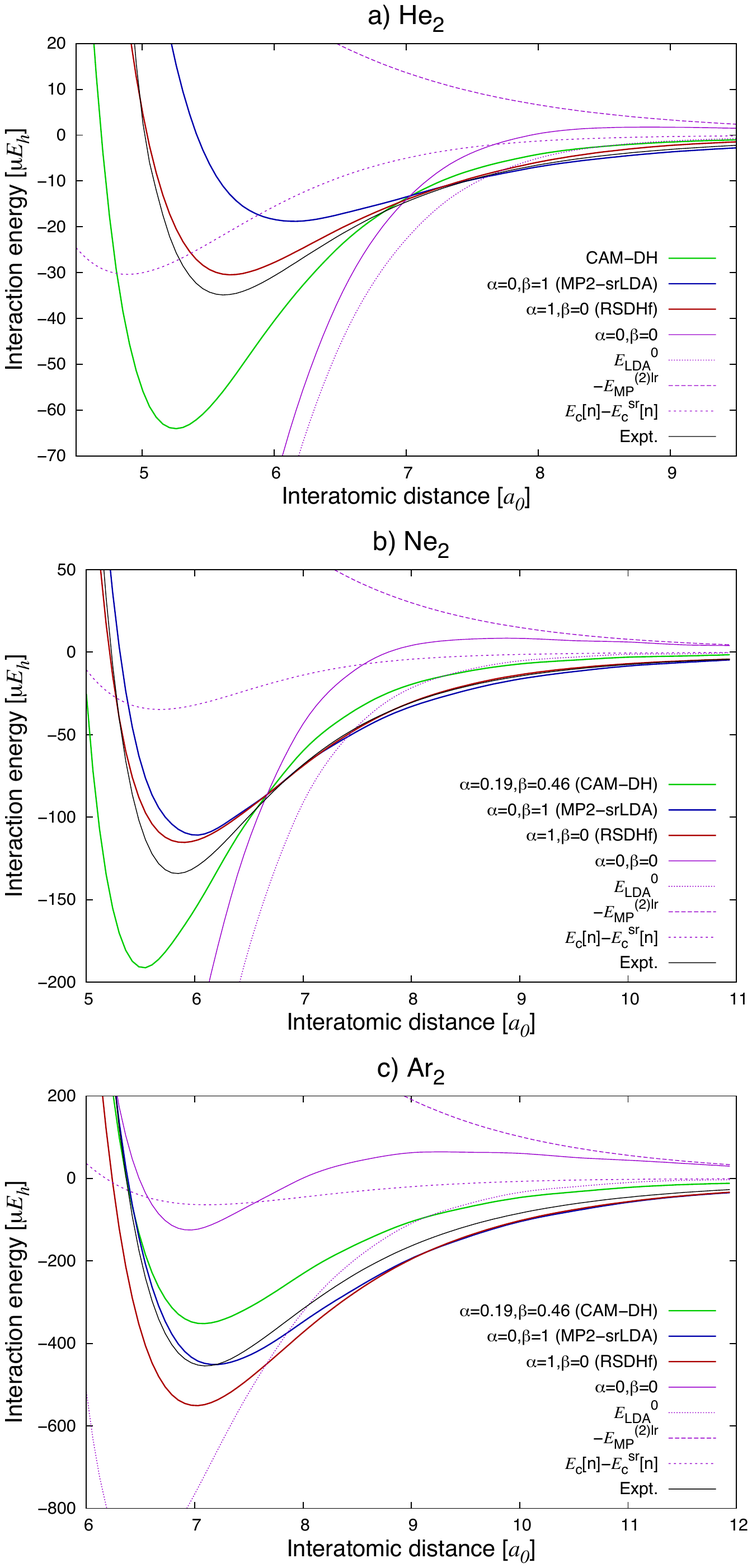}\\
\end{center}
\hspace*{3in}
{\Large
\begin{minipage}[t]{3in}
\baselineskip = .5\baselineskip
Figure 1 \\
Yann Cornaton, Emmanuel Fromager\\
Int. J.\ Quant.\ Chem.
\end{minipage}
}

\clearpage

\begin{center}
\includegraphics[width=0.85\columnwidth,keepaspectratio=true]{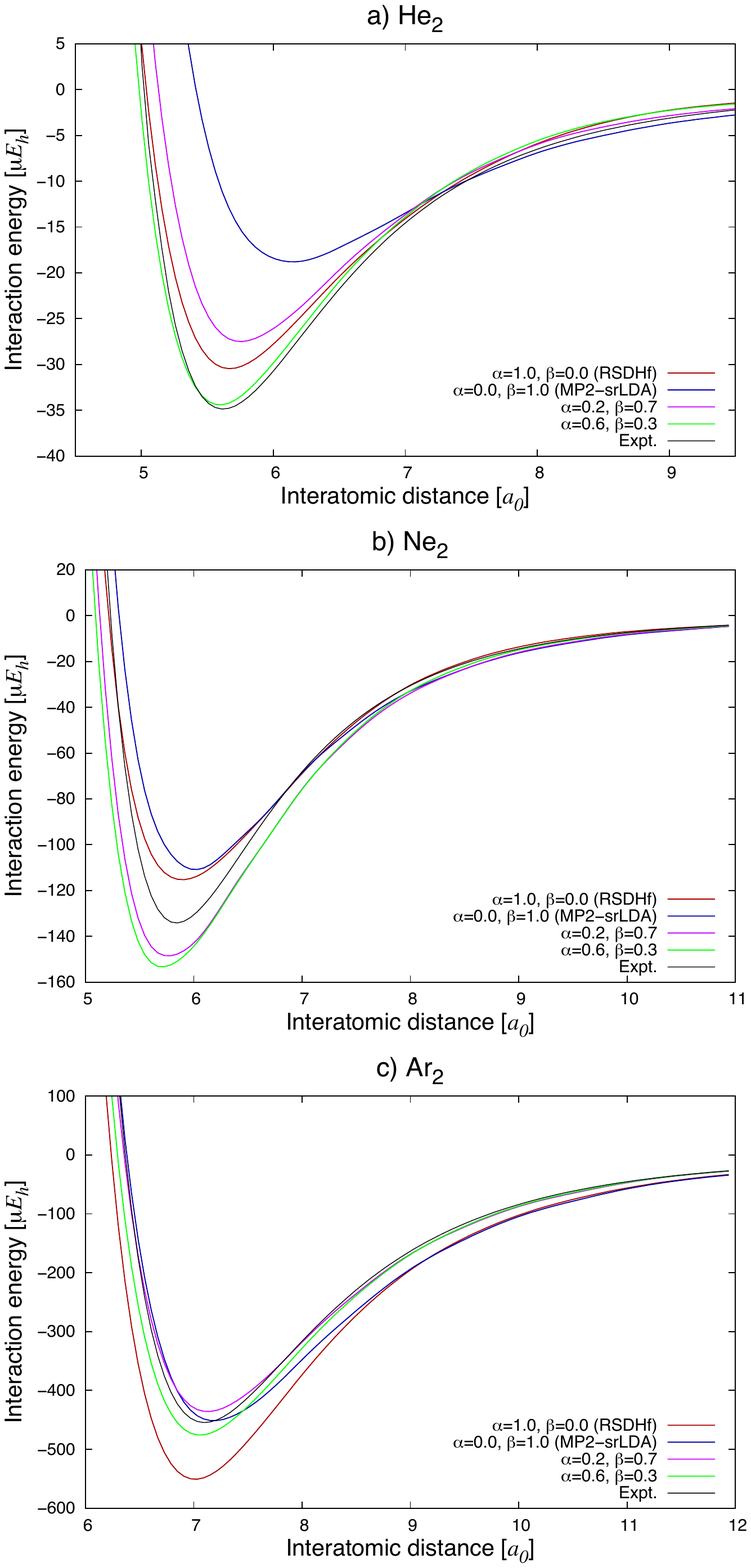}\\
\end{center}
\hspace*{3in}
{\Large
\begin{minipage}[t]{3in}
\baselineskip = .5\baselineskip
Figure 2 \\
Yann Cornaton, Emmanuel Fromager\\
Int. J.\ Quant.\ Chem.
\end{minipage}
}

\clearpage

\begin{center}
\includegraphics[width=0.95\columnwidth,keepaspectratio=true]{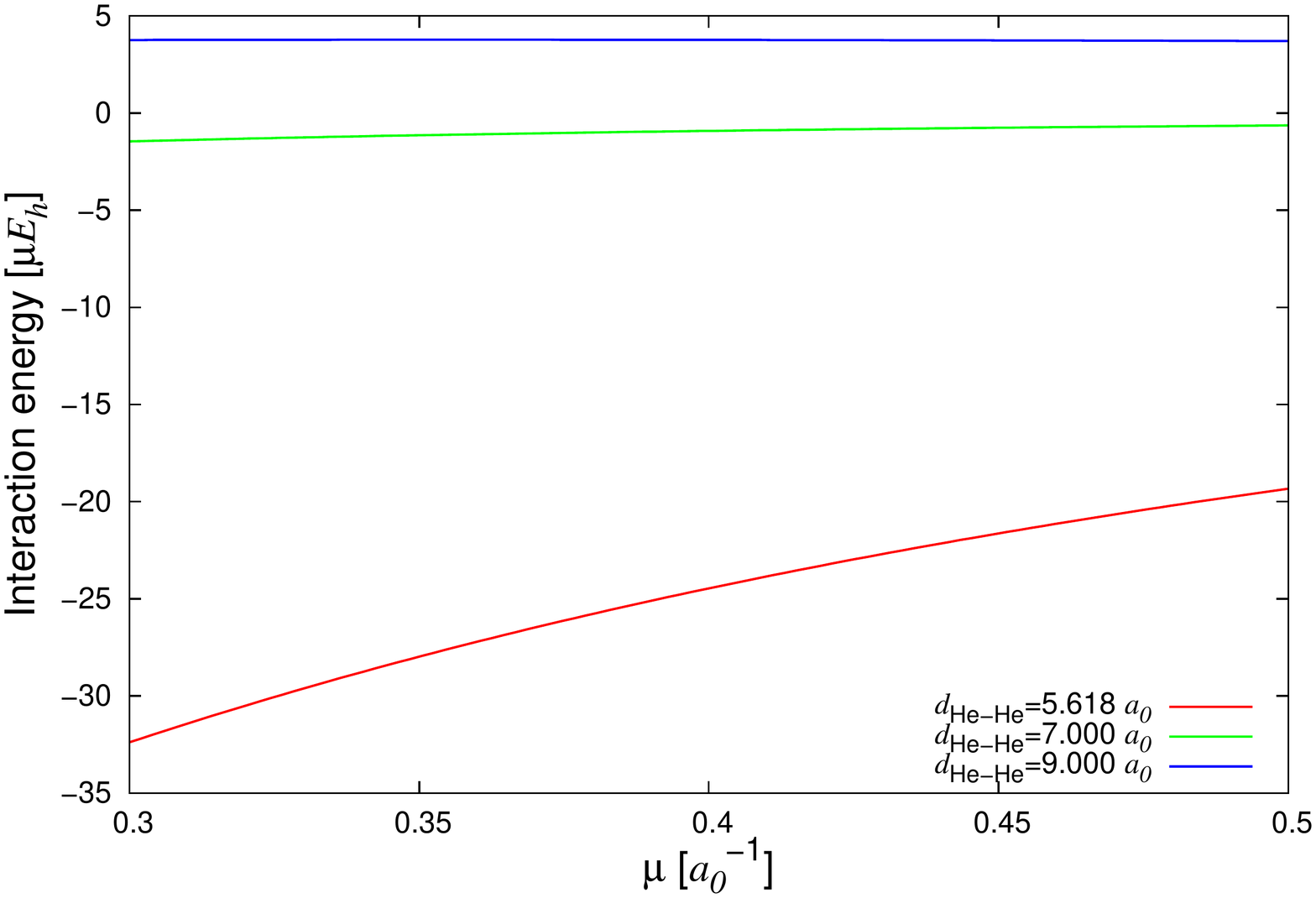}
\end{center}
\vspace{0.25in}
\hspace*{3in}
{\Large
\begin{minipage}[t]{3in}
\baselineskip = .5\baselineskip
Figure 3 \\
Yann Cornaton, Emmanuel Fromager\\
Int. J.\ Quant.\ Chem.
\end{minipage}
}

\clearpage

\begin{center}
\includegraphics[width=0.85\columnwidth,keepaspectratio=true]{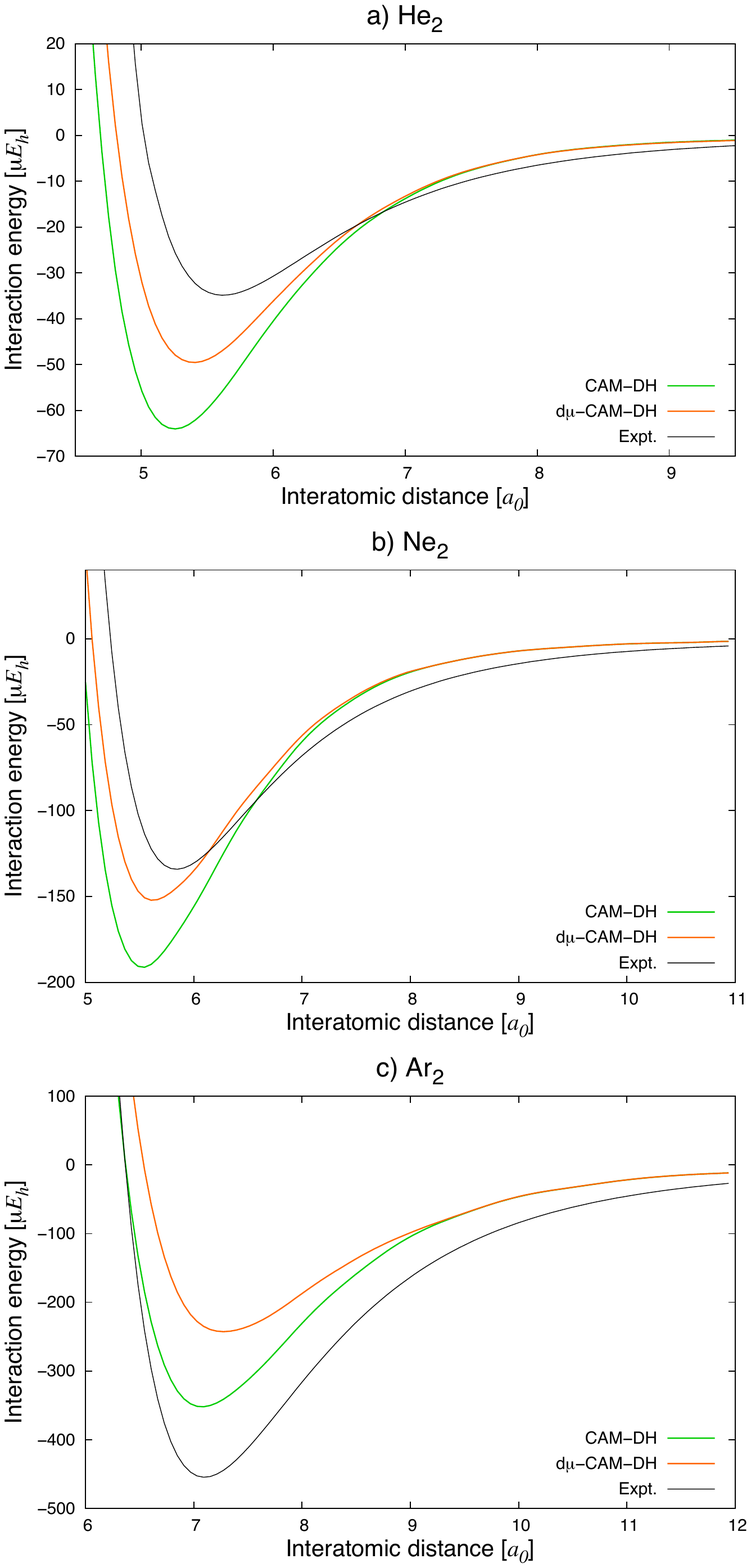}\\
\end{center}
\hspace*{3in}
{\Large
\begin{minipage}[t]{3in}
\baselineskip = .5\baselineskip
Figure 4 \\
Yann Cornaton, Emmanuel Fromager\\
Int. J.\ Quant.\ Chem.
\end{minipage}
}

\clearpage

\begin{center}
\includegraphics[width=0.85\columnwidth,keepaspectratio=true]{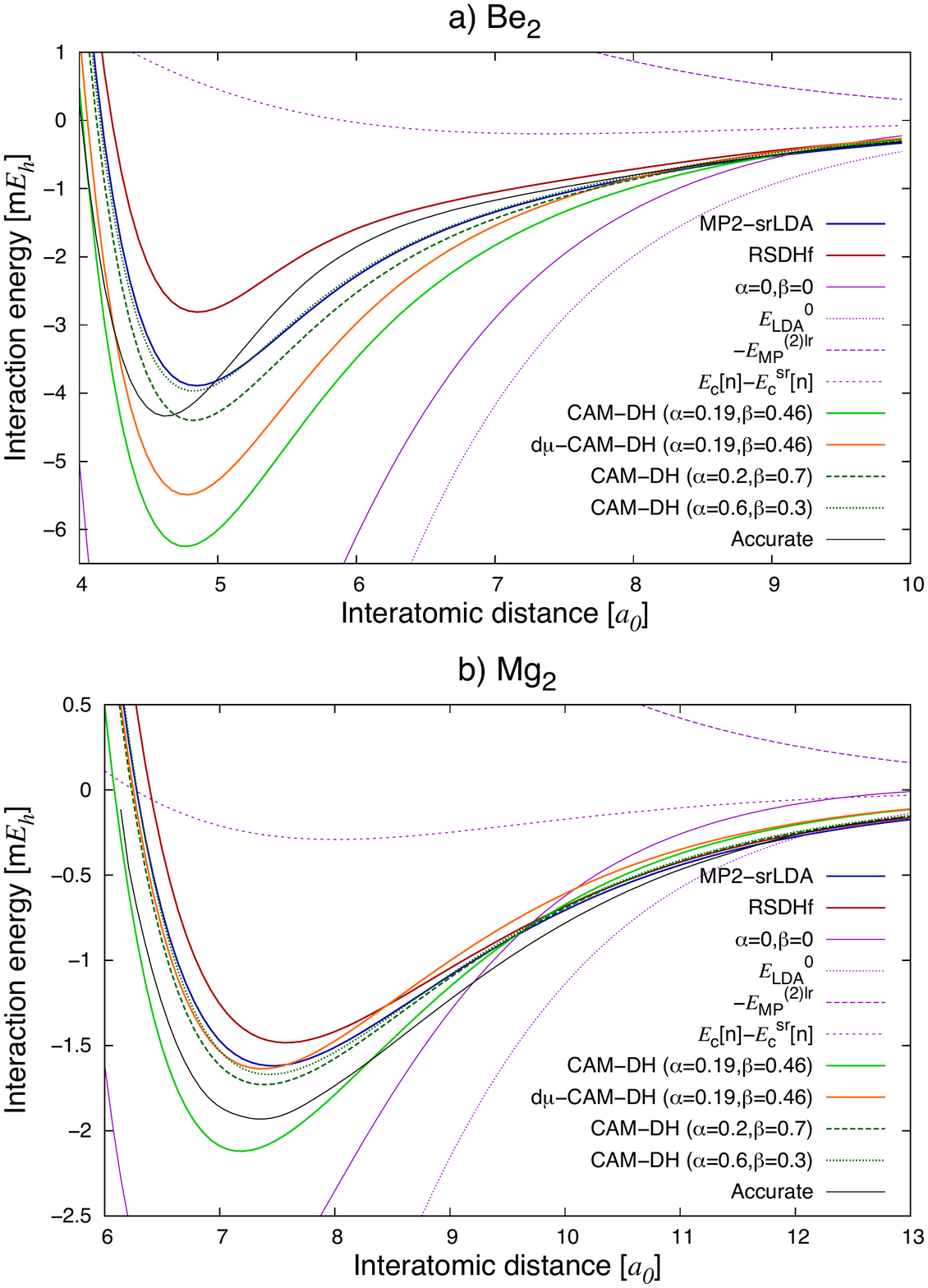}\\
\end{center}
\hspace*{3in}
{\Large
\begin{minipage}[t]{3in}
\baselineskip = .5\baselineskip
Figure 5 \\
Yann Cornaton, Emmanuel Fromager\\
Int. J.\ Quant.\ Chem.
\end{minipage}

\clearpage

\begin{center}
\includegraphics[width=0.85\columnwidth,keepaspectratio=true]{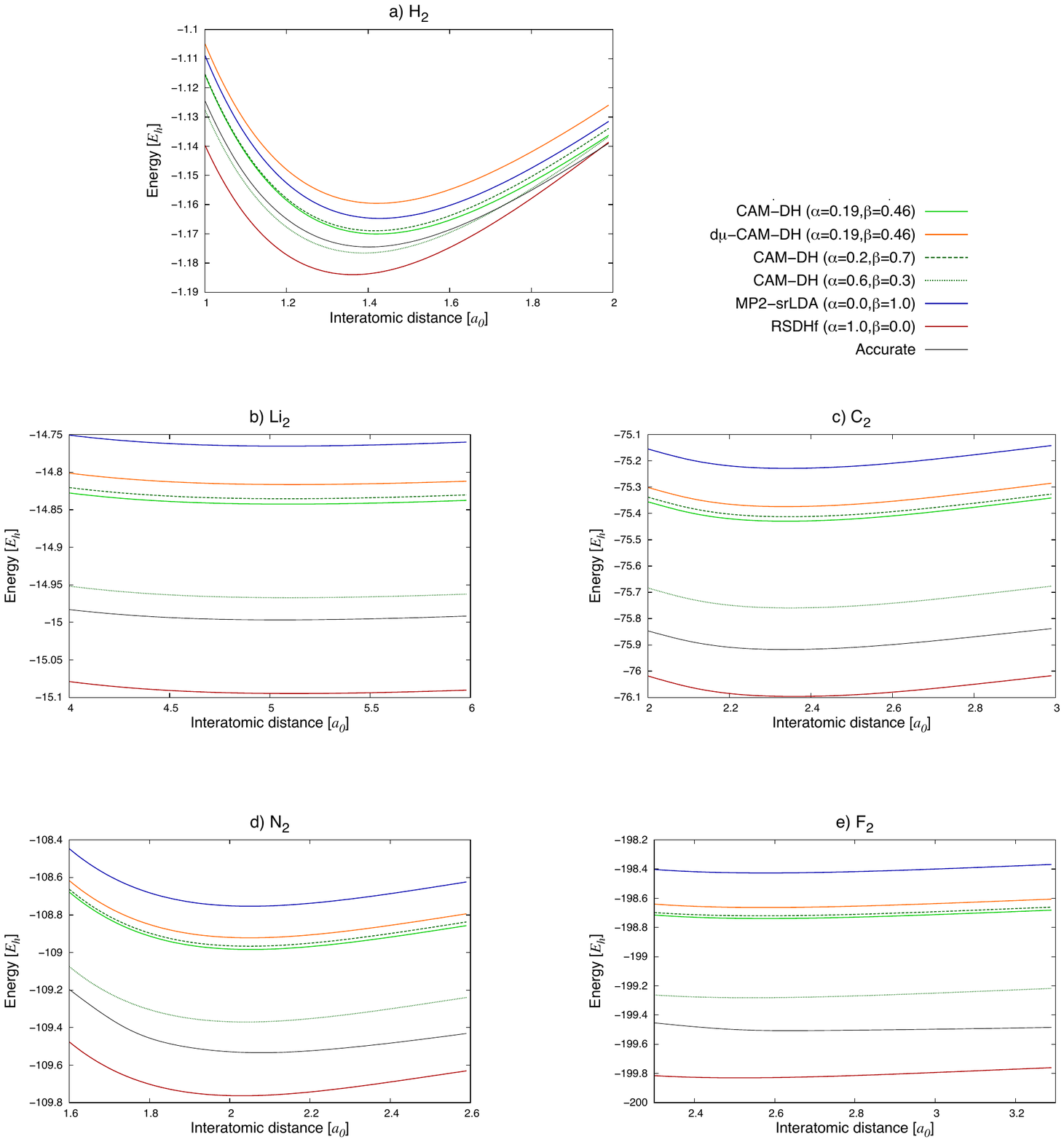}\\
\end{center}
\hspace*{3in}
{\Large
\begin{minipage}[t]{3in}
\baselineskip = .5\baselineskip
Figure 6 \\
Yann Cornaton, Emmanuel Fromager\\
Int. J.\ Quant.\ Chem.
\end{minipage}

}


\end{document}